\documentclass[pre,aps,epsfig,twocolumn,superscriptaddress,showpacs]{revtex4}
\usepackage{graphicx}

\begin{document}
\title{Collective synchronization in spatially extended systems of coupled oscillators 
with random frequencies}

\author{H. Hong}
\affiliation{Department of Physics, Chonbuk National University, Chonju 561-756, Korea}
\affiliation{School of Physics, Korea Institute for Advanced Study, Seoul 130-722, Korea}

\author{Hyunggyu Park}
\affiliation{School of Physics, Korea Institute for Advanced Study, Seoul 130-722, Korea}

\author{M. Y. Choi}
\affiliation{School of Physics, Korea Institute for Advanced Study, Seoul 130-722, Korea}
\affiliation{Department of Physics, Seoul National University, Seoul 151-747, Korea}

\date{\today}
\begin{abstract}
We study collective behavior of locally coupled limit-cycle oscillators with random
intrinsic frequencies, spatially extended over $d$-dimensional hypercubic lattices. 
Phase synchronization as well as frequency entrainment are explored analytically in the linear 
(strong-coupling) regime and numerically in the nonlinear (weak-coupling) regime.  
Our analysis shows that the oscillator phases are always desynchronized up to $d=4$, which 
implies the lower critical dimension $d_{l}^{P}=4$ for phase synchronization. 
On the other hand, the oscillators behave collectively in frequency (phase velocity)
even in three dimensions ($d=3$), indicating that the lower critical dimension for frequency 
entrainment is $d_{l}^{F}=2$.  Nonlinear effects due to periodic nature of limit-cycle 
oscillators are found to become significant in the weak-coupling regime: 
So-called {\em runaway oscillators} destroy the synchronized (ordered) phase and 
there emerges a fully random (disordered) phase. 
Critical behavior near the synchronization transition into the fully random phase is
unveiled via numerical investigation.  Collective behavior of globally-coupled oscillators 
is also examined and compared with that of locally coupled oscillators.
\end{abstract}
\maketitle

\section{Introduction}
Various systems in nature have been known to exhibit  remarkable 
phenomena of collective synchronization, which have attracted much attention in the 
science community.  To understand collective synchronization behavior, 
systems of coupled oscillators have been widely considered.
One of the simplest and typical models for those systems was first 
introduced by Winfree~\cite{ref:Winfree}, and later refined by Kuramoto and 
others~\cite{ref:Kuramoto0,ref:Kuramoto,ref:synch}, 
considering additional ingredients relevant to reality. 
In existing literature systems of fully coupled oscillators have mostly been considered 
due to its analytical tractability and simplicity. 
On the other hand, systems of spatially extended oscillators with local couplings 
have received less attention  even though they are more realistic and 
frequently observed in nature. 
In some studies on the locally-coupled 
oscillators~\cite{ref:Sakaguchi,ref:Daido,ref:Bahiana,ref:Aoyagi,ref:Strogatz,ref:Pikovsky},
collective synchronization, in particular, frequency entrainment has been investigated. 
However, they did not provide a clear answer as yet to whether the ordered (frequency-entrained)
phase exists in space dimension $d=2$, consequently the value of the lower critical dimension,
and also to the critical property near the frequency-entrainment transition in higher dimensions. 
Phase synchronization has also been studied; however, there still remain many fundamental
questions that are not clearly answered, including the lower critical dimension for 
phase synchronization and critical scaling properties. 

Collective synchronization in phase or frequency (phase velocity) of infinitely many 
oscillators may emerge via competition between ferromagnetic-type coupling and scatteredness 
of random intrinsic frequencies. In the strong-coupling limit, 
oscillators prefer to behave collectively, overcoming the randomness of intrinsic 
phase velocities. In the weak-coupling limit, each oscillator tends to be obedient to 
its intrinsic frequency and makes the system disordered in phase and/or frequency. 
In low space dimensions the system with local couplings may even not be partially
ordered (synchronized) for arbitrarily strong coupling strength. 

In this paper, we obtain this dimensional threshold for appearance of the ordered phase, i.e.~the 
lower critical dimension for both phase synchronization and frequency entrainment.
Also unveiled is the nature of the synchronization transitions in higher dimensions. 
In the strong-coupling regime, we linearize the equations of motion for the oscillators 
and investigate analytically their collective behavior. 
An analogy to the surface growth problem helps us to probe the synchronization order parameter. 
In the weak-coupling regime, we numerically integrate the equations of motion and find, 
via careful finite-size-scaling analysis, that the synchronized phase 
and the phase synchronization transition appear only for $d\ge 5$ 
while the frequency entrainment appears for $d\ge 3$.

This paper consists of five sections: Section II introduces the 
system of locally-coupled oscillators on $d$-dimensional hypercubic lattices.  
In Sec.~III, phase synchronization is studied by means of the linear theory and numerical
integration.  
Nonlinear effects are interpreted in terms of runaway oscillators 
and the analogy to the surface landscape is discussed. We also investigate the nature of 
the phase synchronization in five and six dimensions. 
This study of phase synchronization is complementary to our recent work~\cite{ref:shortpaper}. 
Section IV is devoted to the study of the frequency entrainment behavior. 
Discussed are nonlinear effects and complete frequency entrainment as well as 
the nature of the frequency entrainment transition. 
In Sec. V the globally coupled system is examined and compared in detail with the 
locally coupled system.  Finally, a brief summary is given in Sec.~VI.

\section{locally coupled oscillators}

A general description of coupled oscillator systems may be given by
\begin{equation}
\frac{d}{dt}{\bf X}_i = {\bf F}_i ({\bf X}_i) + \sum_j {\bf G}_{ij}({\bf X}_i , {\bf X}_j ),
\end{equation}
where ${\bf X}_i$ is the vector describing the $i$th oscillator and 
${\bf G}_{ij}$ represents the coupling between oscillators $i$ and $j$.
When the coupling is not strong (i.e., for small ${\bf G}$), one can show that only phases
(rather than amplitudes) of oscillators are relevant (see, e.g., Ref. \onlinecite{ref:Kuramoto0}). 
It is well known that, as the coupling becomes sufficiently strong, the amplitude 
variation may not be neglected and one should resort to the complex Ginzburg-Landau (CGL) 
description; this is often suitable for describing the synchronization phenomena of 
identical oscillators, which overcome the frequency variations originated from 
amplitude variations. 

In this paper, we do not intend to cover the regime where the CGL description is valid.
We instead focus on the collective synchronization transition displayed by a large number of 
limit-cycle oscillators with random intrinsic frequencies, 
where only oscillator phases are relevant fluctuating variables. 

We thus consider the set of equations of motion 
\begin{equation}
\frac{d\phi_i}{dt} = \omega_i - K\sum_{j\in \Lambda_i}\sin(\phi_i - \phi_j),
\label{eq:model}
\end{equation}
which governs the dynamics of $N$ coupled limit-cycle oscillators located at sites of 
a $d$-dimensional hypercubic lattice. 
Here $\phi_i$ represents the phase of the $i$th oscillator $(i=1,2,\cdots,N)$, whereas
the first term and the second term on the right-hand side represent the intrinsic frequency of 
the $i$th oscillator and the local interactions between the $i$th oscillator and its 
nearest neighbors, the set of which is denoted by $\Lambda_i$, respectively. 
The intrinsic frequency $\omega_i$ is assumed to be randomly distributed according to 
the Gaussian distribution function $g(\omega)$ with mean $\omega_0$, 
which we set $\omega_0\equiv 0$ without loss of generality, and variance $2\sigma$. 
The coupling is assumed to be {\em ferromagnetic}, i.e., $K>0$, so that 
neighboring oscillators favor their phase difference minimized. 
The sine function form is the most general representation of the coupling in the lowest order 
and its periodic nature is generic in limit-cycle oscillator systems. 
Higher-order terms are irrelevant in the sense of universality. 
It is also noteworthy that Eq.~(\ref{eq:model}) with
spatio-temporal random noise (thermal noise) $\eta_i (t)$ instead of the quenched noise 
$\omega_i$ describes the dynamics of the well-known O(2)-symmetric XY model.

When the coupling is absent ($K=0$), each limit-cycle oscillator evolves with 
its own intrinsic frequency $\omega_i$ according to $d\phi_i/dt=\omega_i$, 
resulting in that the system becomes trivially desynchronized.
For finite coupling $(K> 0)$, locally ordered (synchronized) regions emerge,
inside of which oscillators evolve with a coupling-modified effective frequency. 
Here the dispersion of intrinsic frequencies competes against the coupling 
and this competition sets the size of locally ordered regions. 
When the coupling is strong enough to overcome the dispersion of intrinsic frequencies
and subsequently to create a globally ordered region, the system exhibits collective 
synchronization behavior. 

For the system of coupled limit-cycle oscillators, 
two kinds of collective synchronization may be considered:
frequency synchronization and phase synchronization.  
The former is often called frequency entrainment or {\em phase locking}. 

\section{phase synchronization}

We first investigate the phase synchronization, which may be probed by 
the conventional phase order parameter
\begin{equation}
\Delta\equiv \left\langle \frac{1}{N}\left|\sum_{j=1}^{N}e^{i\phi_j}\right|\right\rangle
\label{eq:def_order}
\end{equation}
with $\langle\cdots\rangle$ denoting the average over different realizations
of intrinsic frequencies.  A nonzero value of $\Delta$ then implies the emergence of 
phase synchronization. 

Up to date, analytic treatment has been available only at the mean-field (MF) 
level (see Sec.~VII).  Namely, in the case of globally coupled oscillators where
each oscillator is coupled to every other one with equal strength $K/N$, 
it is known that the system exhibits collective synchronization in phase, 
which is described by $\Delta\sim (K-K_c)^{\beta}$ with $\beta=1/2$ near the critical 
coupling strength $K_c=2/\pi g(0)$~\cite{ref:Kuramoto}. 
It is of interest to note that both the phase synchronization and the frequency entrainment 
emerge simultaneously at the same coupling strength in the globally coupled system. 

\subsection{Linear theory}

Systems with locally coupled oscillators have been little investigated, 
presumably due to the difficulty in analytical treatment. 
The nonlinear nature of the sine coupling term in Eq.~(\ref{eq:model}) is the major 
obstacle toward analytic treatment. 
Accordingly, we first suppose that, for sufficiently strong coupling strength $K$, 
the phase difference between any nearest neighboring oscillators is small enough 
to allow the expansion of the sine function in the linear regime.  
Taking the appropriate continuum limit in space, we obtain Eq.~(\ref{eq:model}) in the form 
\begin{equation}
\frac{\partial}{\partial t}\phi({\bf x},t)
= \omega({\bf x}) + K\nabla^2 \phi
+ {\cal {O}}\left(\nabla^4 \phi, (\nabla\phi)^2\nabla^2\phi  \right),
\label{eq:linear}
\end{equation}
where $\omega ({\bf x})$ are uncorrelated random variables, satisfying 
$\langle\omega({\bf x})\rangle=0$ and 
$\langle\omega({\bf x})\omega({\bf x^\prime})\rangle = 2\sigma\delta({\bf x}-{\bf x^\prime})$. 
We note that the coupling constant $K$ and the variance $2\sigma$ may be renormalized 
through a coarse graining procedure in taking the continuum limit.  However, the 
renormalizing factor may be absorbed into the time scale via proper rescaling of time. 
For convenience, we also relax the constraint $0 \le \phi < 2\pi$ and extend the range to 
$(-\infty , \infty)$, which is permissible in the linear regime. 

Equation~(\ref{eq:linear}), with the irrelevant higher-order terms neglected, 
is reminiscent of the well-known Edwards-Wilkinson (EW) equation~\cite{ref:EW}. 
The EW equation traditionally describing surface evolution becomes identical
to Eq.~(\ref{eq:linear}) by interpreting the phase $\phi({\bf x},t)$ as the front height of 
the growing surface. 
Note, however, that the noise $\omega({\bf x})$ is generated not by 
conventional spatio-temporal disorder but by so-called columnar disorder which
has only spatial dependence. In other words, the disorder $\omega({\bf{x}})$ is
independent of time, which is different from the conventional thermal noise.
The coupling strength $K$ plays the role of surface tension in the growing surface model. 

A key quantity of interest in the context of surface growth models 
is the surface fluctuation width $W$ defined by 
\begin{equation}
W^2(t) = \frac{1}{L^d}\int^{L} d^d x
\left\langle \left[ \phi({\bf x},t)- {\bar \phi}(t) \right]^2 \right\rangle,
\label{eq:W}
\end{equation}
where $L$ is the linear size of the $d$-dimensional lattice $(L^d = N)$ and 
${\bar \phi}(t)$ is the spatial average
\begin{equation}
\bar \phi(t) \equiv \frac{1}{L^d} \int^{L} d^d x\, \phi({\bf{x}},t).
\end{equation}
Since Eq.~(\ref{eq:linear}) is linear, it is convenient to consider the Fourier transform
\begin{equation}
\phi({\bf{x}},t)=\frac{1}{(2\pi)^d}\int d^d k\, \phi({\bf{k}},t)
e^{i{\bf{k}}\cdot {\bf{x}}},
\end{equation}
in terms of which Eq.~(\ref{eq:linear}) reads 
\begin{equation}
\frac{\partial}{\partial t}\phi({\bf{k}},t)=\omega({\bf{k}})-Kk^2 \phi({\bf{k}},t)
\label{eq:linear_k}
\end{equation}
with higher-order terms neglected. 
The solution of Eq.~(\ref{eq:linear_k}) is easily obtained as 
\begin{equation}
\phi({\bf{k}},t)=\phi({\bf{k}},0)e^{-Kk^2 t}+
\frac{\omega({\bf{k}})}{Kk^2} (1-e^{-Kk^2 t}),
\label{eq:solution}
\end{equation}
which in turn gives the mean-square width
\begin{eqnarray}
W^2 &=& \frac{1}{(2\pi)^{2d}} \int{d^d k}
\int d^d{k^{'}} \langle \phi({\bf{k}},t)\phi({\bf{k^{'}}},t) \rangle \nonumber\\
&=& \frac{1}{(2\pi)^{2d}}\int d^d{k} \int {d^d{k^{'}}}
\frac{2\sigma}{K^2k^2{k^{'}}^2}
(2\pi)^d \delta^d({\bf{k}}+{\bf{k^{'}}})\nonumber\\
& & ~~~\times (1-e^{-Kk^2 t})(1-e^{-K{k^{'}}^2 t}) \nonumber\\
&=& \frac{2\sigma\Omega_d}{K^2}\int_{2\pi/L}^{\pi/a} dk k^{d-5} 
(1-2e^{-Kk^2 t}+ e^{-2Kk^2 t})
\label{eq:W2}
\end{eqnarray}
with $\Omega_d \equiv 2^{1-d}\pi^{-d/2}/\Gamma(d/2)$ and $a$ denoting the lattice constant. 
In obtaining Eq.~(\ref{eq:W2}), we have used the relation
$\langle\omega({\bf{k}})\omega({\bf{k^{'}}})\rangle=2\sigma(2\pi)^d 
\delta^d ({\bf{k}}+{\bf{k^{'}}})$ and taken $\phi({\bf{k}},0)=0$ for the initial condition. 

In the long-time limit $(Kt\gg L^2)$, the surface width in the stationary state 
scales, for large $L$:
\begin{eqnarray}
\label{eq:W2_result}
W^2  &\sim& ( 2\sigma/ K^2) L^{4-d},  ~~~~~~d <4 \nonumber\\
     &\simeq& (\sigma/4\pi^2 K^2) \ln L ,~~~~d =4 \\
     &\sim&   2\sigma/K^2 ,     ~~~~~~~~~~~~~~~d>4. \nonumber
\end{eqnarray}
At any finite values of $K$, the surface width $W$ thus diverges as $L\rightarrow \infty$
for $d\le 4$ whereas it remains finite for $d> 4$.  This indicates that the surface
is always rough (except at $K=\infty$) for $d\le 4$ and
always smooth (except at $K=0$) for $d>4$.
In the short-time regime ($Kt \ll L^2$), $W^2$ in Eq.~(\ref{eq:W2}) becomes 
\begin{eqnarray}
\label{eq:W2_result2}
W^2  &\sim& ( 2\sigma/ z K^2) (Kt)^{(4-d)/z},  ~~~~~~d <4  \nonumber \\
     &\simeq& (\sigma/4\pi^2 z K^2) \ln (Kt) ,~~~~~~~~~d =4,
\end{eqnarray}
with the dynamic exponent $z=2$.  
For $d>4$, the mean-square width $W^2$ thus saturates to $\sim 2\sigma/K^2$ exponentially. 

As expected, the exponents $\alpha$ and $\beta$, which characterize the width fluctuations 
according to $W\sim L^{\alpha}$ and $W\sim t^{\beta}$ in the long- and short-time limits, 
respectively, are different from those of the original EW model with conventional thermal noise. 
The difference is attributed to the quenched columnar noise which does not have time dependence. 
In fact, simple power counting yields the values of $\alpha$ and $\beta$, 
which are consistent with Eqs.~(\ref{eq:W2_result}) and (\ref{eq:W2_result2}). 

We can also derive analytically the exact stationary-state probability distribution 
$P[\{\phi\}]$ in the linear regime described by Eq.~(\ref{eq:linear}). 
Note that it is usually quite difficult to find the exact distribution function 
for a system with quenched disorder.  Equation~(\ref{eq:linear_k}) assures that 
$\phi_{\bf{k}}$ should become $\omega_{\bf{k}}/(K{\bf{k}}^2)$ for any $\bf{k}$ in the 
stationary state, where $\phi_{\bf{k}}$ and $\omega_{\bf{k}}$ are used in place of 
$\phi(\bf{k})$ and $\omega(\bf{k})$, respectively, for brevity. 
Accordingly, one can write the stationary-state probability of finding the configuration 
$\{\phi_{\bf{k}}\}\equiv \{\phi_{\bf{k_1}},\cdots,\phi_{\bf{k_N}}\}$ 
for given distribution $\{\omega_{{\bf{k}}}\}$:
\begin{equation}
P_{\omega_{{\bf{k}}}}(\{\phi_{\bf{k}}\})=\prod_{\bf{i}}\delta\left(\phi_{\bf{k_i}}-
\frac{\omega_{\bf{k_i}}}{K{\bf{k_i}}^2}\right).
\label{eq:prob}
\end{equation}
Averaging over random frequencies $\{\omega_{{\bf{k}}}\}$, we find 
\begin{eqnarray}
&&P[\{\phi\}] \sim \int {\cal{D}}{\omega_{{\bf{k}}}} 
P_{\omega_{{\bf{k}}}}(\{\phi_{\bf{k}}\}) \nonumber\\
&&= \int d\omega_{{\bf{k_1}}} \cdots d\omega_{{\bf{k_N}}}
\exp \left[-\sum_{\bf i} \frac {\omega^2_{\bf{k_i}}}{4\sigma}\right] 
\prod_{\bf{i}}\delta\left(\phi_{\bf{k_i}}-
\frac{\omega_{\bf{k_i}}}{K{\bf{k_i}}^2}\right) \nonumber\\
&&=\exp \left[-\sum_{\bf i} \frac {K^2}{4\sigma}
{\bf{k^4_i}}\phi^2_{{\bf{k_i}}} \right]  \nonumber\\
&&= \exp \left[ -\frac {K^2}{4\sigma}\int \left(\nabla^2\phi \right)^2 d\bf{x}\right].
\label{eq:Gaussian}
\end{eqnarray}
The mean-square width $W^2$ in the stationary state, given by Eq.~(\ref{eq:W2_result}), 
may be derived directly from this distribution function. 

It is of much interest to note that this nonequilibrium stationary-state probability 
distribution is identical to that of the equilibrium 
{\em Laplacian (tensionless) roughening model}~\cite{ref:Laplacian}. 
The prefactor $K^2/(4\sigma)$ in the exponent of Eq.~(\ref{eq:Gaussian}) plays the
role of the inverse temperature in the Laplacian roughening model. 
One may also relate our model to the {\em linear molecular beam epitaxy (MBE) growth} 
model~\cite{ref:DasSarma}, the evolution dynamics of which is described by
\begin{equation}
\frac{\partial}{\partial t}\phi({\bf x},t)
= \eta({\bf x},t) - \nu\nabla^4 \phi,
\label{eq:MBE}
\end{equation}
where $\eta({\bf x},t)$ is the spatio-temporal random noise satisfying
$\langle\eta({\bf x},t)\rangle=0$ and 
$\langle\eta({\bf x},t)\eta({\bf x^\prime},t^\prime)\rangle 
= 2D \delta({\bf x}-{\bf x^\prime})\delta(t-t^\prime)$. 
It is straightforward to show that the stationary distribution of this MBE model
is also identical to Eq.~(\ref{eq:Gaussian}) with the correspondence 
$D\leftrightarrow\sigma$ and $\nu\leftrightarrow K^2/2$. 
However, the dynamic behavior of the MBE model is characterized by the dynamic exponent $z=4$, 
in contrast to $z=2$ in our model. 

Another important observation is the Gaussian property of the stationary distribution,
which provides a link between the mean-square width $W^2$ and the phase order 
parameter $\Delta$.
The practical definition of $\Delta$ in Eq.~(\ref{eq:def_order}) is not convenient 
for analytical treatment, and instead we use the formal but equivalent definition:
$\Delta\equiv \langle e^{i(\phi-{\bar \phi})}\rangle$, where ${\bar\phi}$ denotes
the spatial average. The Gaussian property of the stationary probability 
distribution in Eq.~(\ref{eq:Gaussian}) guarantees 
$\langle e^{i f(\phi)}\rangle=e^{-\langle f^2(\phi)\rangle/2}$ 
for an arbitrary function $f(\phi)$. 
Therefore, one can easily obtain 
\begin{equation}
\label{eq:Delta-W}
\Delta=\exp \left[ -W^2/2 \right]
\end{equation}
in the stationary state.  It should be noted that this relation is valid only 
in the linear (strong-coupling) regime described by Eq.~(\ref{eq:linear}). 

With this relation, we can express our results for the width $W$ in the phase 
synchronization language: The oscillators are always desynchronized (disordered; $\Delta=0$) 
for $d\le 4$ in the thermodynamic limit ($N\rightarrow\infty$), whereas
they are always partially synchronized (ordered; $\Delta\neq 0$) for $d>4$ 
at any finite coupling, leading to $\Delta=\exp (-A_d \sigma/K^2)$ with 
$A_d= \Omega_d \pi^{d-4} /(d-4)$. Accordingly, in the framework of the linear theory, 
there is neither phase synchronization-desynchronization transition nor 
complete phase synchronization ($\Delta=1$) at any finite coupling $K$ in any dimension $d$. 

Our linear theory is valid in the strong-coupling regime; as the weak coupling regime
is approached, the original (nonlinear) system described by Eq.~(\ref{eq:model}) 
should be more disordered than the prediction of the linear theory.
This establishes that the full nonlinear system should also be desynchronized ($\Delta=0$)
for $d\le 4$ at any finite $K$, which implies that the lower critical dimension 
for the phase synchronization may not be less than four: $d_l^P\ge 4$. 
The nature of the desynchronized phase may become different from what 
is expected from the linear theory, especially in the weak-coupling regime. 
At small $K$, the system becomes far more disordered so that the phase difference 
between nearest neighboring oscillators can grow large enough to invalidate 
the expansion of the nonlinear sine function exercised in obtaining the linear theory 
in Eq. (\ref{eq:linear}). 
For $d>4$, it is reasonable to expect a phase synchronization  transition 
(roughening transition in the surface growth language) at a finite value of $K$.
Emergence of the desynchronized phase at nonzero $K$ for $d>4$ should be attributed 
solely to nonlinear effects. Of course, one may not rule out the possibility of
either the full destruction of the synchronized phase at any finite $K$ or
the absence of the desynchronized phase at all. 

Before investigating the full nonlinear system described by 
Eq.~(\ref{eq:model}), we examine the self-consistency of our linear theory
by considering another standard quantity
in surface growth models, the height-height correlation function
\begin{equation}
C({\bf{x}},t)\equiv \left\langle\left[\phi({\bf{x}},t)-\phi({\bf{0}},t)
\right]^2\right\rangle.
\end{equation}
Similar to the case of $W^2$, the correlation function is easily obtained 
by means of the Fourier transform:
\begin{eqnarray}
C({\bf{x}},t)&=&\frac{1}{(2\pi)^d} \int d^d k \,
  \frac{4\sigma}{K^2 k^4}\left(1-2e^{-Kk^2 t}+e^{-2Kk^2 t}\right) \nonumber\\
 & &~~~~~~~~~~~~~~~~~\times\left[1-\cos({{\bf{k}}\cdot{\bf{x}}})\right].
\end{eqnarray} 
For small $x\, (\equiv |\bf{x}|)$, the term $\cos({\bf{k}}\cdot {\bf{x}})$ may be 
expanded as $1-k^2 x^2/2 +{\cal {O}}(x^4)$, which finally yields 
the following stationary behavior: 
\begin{eqnarray}
\label{eq:C_result}
C(x)  &\sim& ( 2\sigma/ K^2) x^2 L^{2-d},  ~~~~~~d <2 \nonumber\\
      &\simeq& (\sigma/2\pi K^2) x^2 \ln L ,~~~~~~d =2 \\
      &\sim&   (2\sigma/ K^2) x^{4-d} ,     ~~~~~~~~~~d>2. \nonumber
\end{eqnarray}
In the short-time regime ($Kt \ll L^2$), on the other hand, we have 
\begin{eqnarray}
\label{eq:C_result2}
C({\bf{x}},t) &\sim& ( 2\sigma/ zK^2) x^2 (Kt)^{(2-d)/z}, ~~~~d <2 \nonumber\\
&\simeq& (\sigma/2\pi zK^2) x^2 \ln (Kt) ,~~~~~~~d =2
\end{eqnarray} 
with $z=2$, and expect exponential saturation for $d>2$. 

Note that for $d\le 2$ the correlation $C({\bf x},t)$ diverges indefinitely with time 
in the thermodynamic limit ($L\rightarrow\infty$). 
With ${\bf x}$ taken as a lattice unit vector, the correlation function represents 
mean-square phase difference between nearest neighboring oscillators. 
For later use, we define the mean-square nearest-neighbor phase difference (step height) 
averaged over all lattice directions as 
\begin{equation}
G^2(K,t)\equiv \left\langle \left[\nabla\phi\right]^2\right\rangle.
\label{eq:G2}
\end{equation}

For $d\le 2$, the average nearest neighbor phase difference $G(K)$ in the stationary state
is unbounded for any finite $K$ in the thermodynamic limit. 
Since our linear theory is based on the boundedness of $|\nabla\phi|$ (in expanding
the nonlinear sine function and dropping off higher-order terms), this implies that 
for $d\le 2$ there is no range of $K$ where the linear theory applies. 
Therefore the nature of the desynchronized phase may be characterized 
not by continuous surface landscape expected from the linear theory 
but possibly by ruptured and splitted surface landscape, which will be discussed later. 
In contrast, for $d> 2$, $G(K)$ remains finite even in the stationary state 
and the linear theory is self-consistent at least for large $K$ where 
$G(K)\lesssim {\cal {O}}(1)$.  For large $G(K)$, the linear theory collapses again, 
which may give rise to the desynchronized phase of discontinuous surface character. 

\subsection{Nonlinear regime}

We now investigate the nonlinear effects which appear due to the sine coupling 
in Eq.~(\ref{eq:model}). In particular, it would be most interesting to probe the possibility 
of emergence of the desynchronized phase at finite coupling strength ($K \neq 0$)
in higher dimensions ($d\ge 5$) and the nature of the synchronization-desynchronization 
transition. In addition, it would also be of interest to understand the nature of the 
desynchronized phase in lower dimensions ($d\le 4$) and possibly a rupturing-type 
phase transition. 

Unlike in the linearized case, the phase difference $|\nabla\phi|$ may not be bounded 
even in a finite system but diverge eventually with finite angular velocity. 
(Recall that in the linear theory the phase difference is always bounded by the 
stationary value of $G$ which is finite for finite system size $L$.) 
Once the intrinsic frequency difference between neighboring oscillators is large enough to
win over the ferromagnetic coupling, the phase difference may grow linearly with time,
indefinitely even in a finite system. 

\begin{figure}
\centering{\resizebox*{!}{6.0cm}{\includegraphics{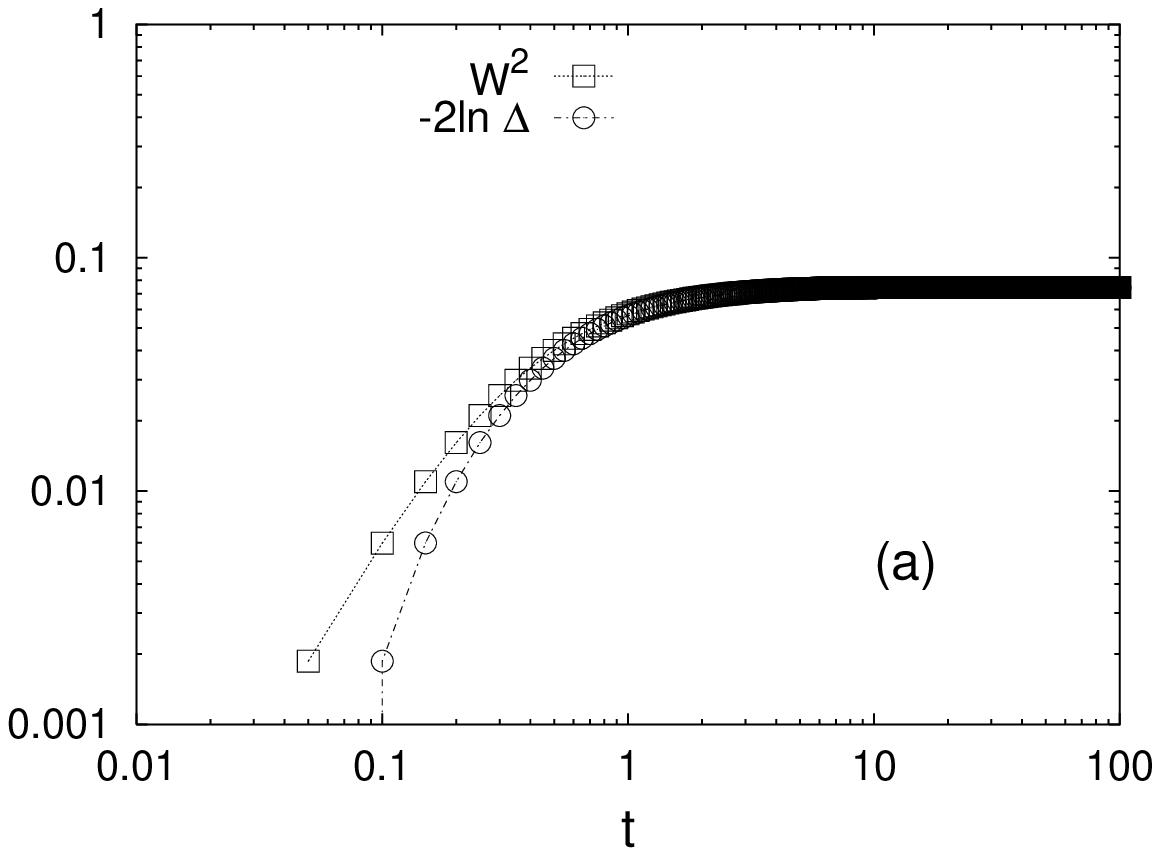}}}
\centering{\resizebox*{!}{6.0cm}{\includegraphics{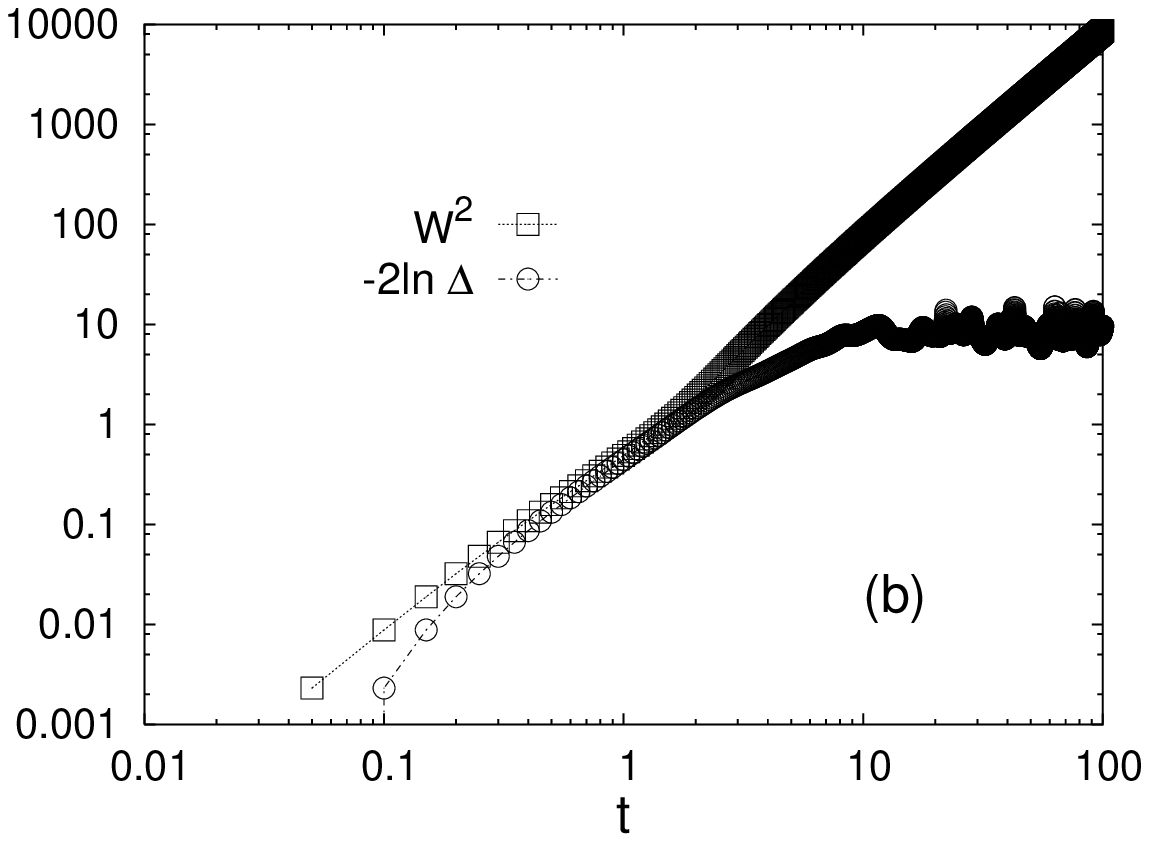}}}
\caption{Phase order parameter $\Delta$ and the mean-square width $W^2$ for 
(a) $K=0.5$ and (b) $K= 0.1$ in $d=5$.}
\label{fig:Delta_W2}
\end{figure}

In the regime of weak coupling, those {\em runaway} oscillators with scattered angular 
velocities dominate and their phases become completely random one another. 
It is easy to show that the phase order parameter defined in Eq.~(\ref{eq:def_order})
should decay algebraically as $\Delta\sim N^{-1/2} = L^{-d/2}$ when the phases $\{\phi_j\}$ 
of $N$ oscillators take fully random values. 
On the other hand, in the strong-coupling regime, where the linear theory applies, 
$\Delta$ vanishes exponentially as $\Delta\sim \exp [-(\sigma/4\pi^3 K^2) L]$ for $d=3$ 
and algebraically as $\Delta \sim L^{-\sigma/(8\pi^2K^2)}$ for $d=4$ 
[see Eqs. (\ref{eq:W2_result}) and (\ref{eq:Delta-W})]. The oscillator phases in this regime 
are desynchronized ($\Delta=0$) as $L\rightarrow\infty$, but they are correlated. 
The landscape of these phases exhibits a continuous surface even if the characteristic 
width diverges with the system size.  In the regime dominated by runaway oscillators, 
the landscape should be very spiky with diverging width, even in a finite system. 
We call this regime the {\em fully random} (desynchronized) phase, while the regime 
where the linear theory applies is dubbed the {\em correlated random} (desynchronized) phase. 
For $d=3$ and $4$, one may expect the transition between the fully random and the 
correlated random phases. Details of this transition will be discussed elsewhere. 

We check numerically the presence of these runaway oscillators in the weak-coupling regime. 
We integrate numerically the full nonlinear equation (\ref{eq:model}) and measure 
both $\Delta$ in Eq.~(\ref{eq:def_order}) and $W^2$ in Eq.~(\ref{eq:W2}). 
In the linear regime, $\Delta$ and $W^2$ should satisfy the relation in Eq.~(\ref{eq:Delta-W})
in the stationary state. On the other hand, in the nonlinear regime with runaway oscillators, 
Eq.~(\ref{eq:Delta-W}) is no longer valid and $W$ is expected to grow linearly with time, 
without saturation while the order parameter $\Delta$ should saturate in a finite system. 

In Fig.~\ref{fig:Delta_W2} we plot the time dependence of the mean-square width $W^2$ 
and $-2\ln \Delta$ at two different coupling strengths ($K=0.5$ and $0.1$) in $d=5$. 
Figure~\ref{fig:Delta_W2}(a) manifests that for the strong coupling ($K=0.5$) 
the relation in Eq.~(\ref{eq:Delta-W}) is well satisfied in the saturated regime 
(i.e., in the stationary state). 
On the other hand, when the coupling is weak ($K=0.1$), the breakdown of the relation 
is evident in Fig.~\ref{fig:Delta_W2}(b). 
In addition, we find that $W$ grows linearly with time $t$ in the long-time limit. 
This linear growth starts rather randomly in time, depending on the disorder realization
(i.e., the distribution of random intrinsic frequencies) and also on the initial condition. 
Averaging over the disorder realization and over the initial condition, 
we still obtain the linear growth of the width $W$. 
Finite-size dependence $\Delta\sim N^{-1/2}$ also supports the presence 
of runaway oscillators, which will be discussed in the next subsection. 

\subsection{Numerical results}

In this subsection, we explore collective phase synchronization of the coupled oscillators 
described by Eq.~(\ref{eq:model}).  We integrate numerically  Eq.~(\ref{eq:model}) and 
measure the phase order parameter $\Delta$ at various values of the coupling strength $K$ and 
the system size $L$.  For convenience, the Gaussian distribution with unit variance 
($2\sigma=1$) has been chosen for the distribution of intrinsic frequencies, 
$g(\omega)\sim \exp (-\omega^2/4\sigma)$, and periodic boundary conditions are employed. 
We begin with the uniform initial condition ($\phi_i = 0$) for given set of 
$\{\omega_i\}$, chosen randomly according to $g(\omega)$. 
We then use Heun's method~\cite{ref:Heun} to integrate Eq.~(\ref{eq:model})
with various values of the discrete time increment $\delta t$. 
Typically, the integration has been performed up to $N_t=4\times 10^4$ time steps with 
$\delta t = 0.05$. We measure the order parameter $\Delta$ averaged over the data 
in the stationary state, reached after appropriate transient time ($Kt\gg L^2$).  
For this purpose, we have discarded data typically to the first $2.8\times 10^4$ time steps. 
Both $\delta t$ and $N_t$ have been varied to check possible systematic deviations;
it has been confirmed that so such deviations were observed within statistical errors. 
For the disorder average over the distribution $g(\omega)$, we have 
carried out one hundred independent runs with randomly chosen realizations $\{\omega_i\}$, 
over which the averages are taken. 

\begin{figure}
\centering{\resizebox*{!}{6.0cm}{\includegraphics{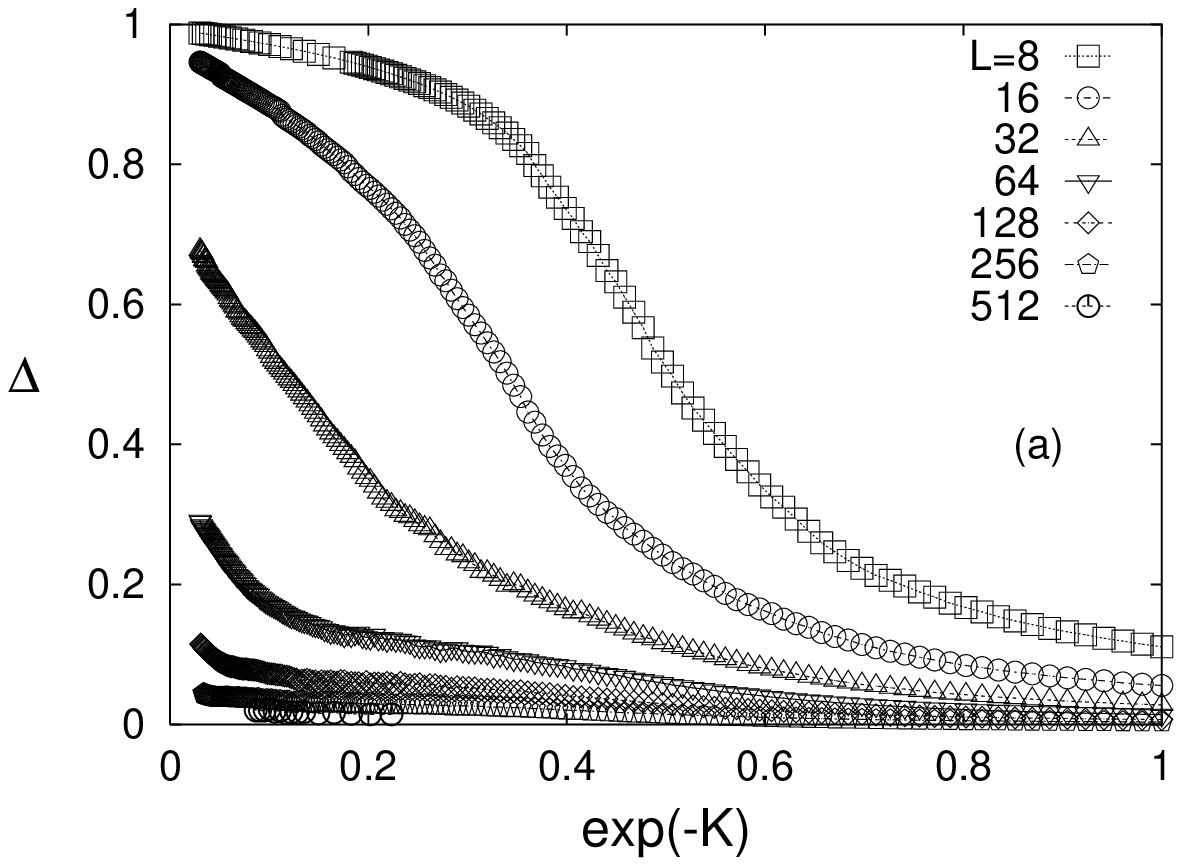}}}
\centering{\resizebox*{!}{6.0cm}{\includegraphics{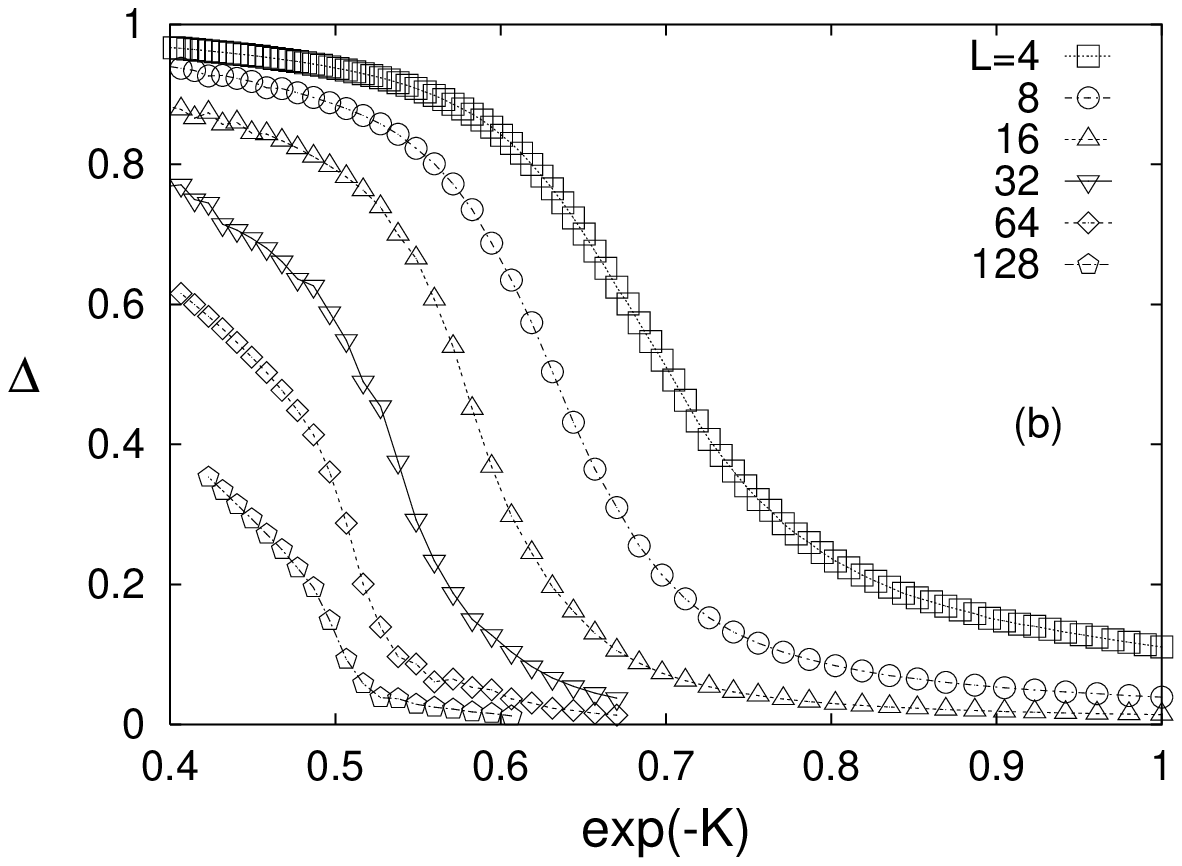}}}
\centering{\resizebox*{!}{6.0cm}{\includegraphics{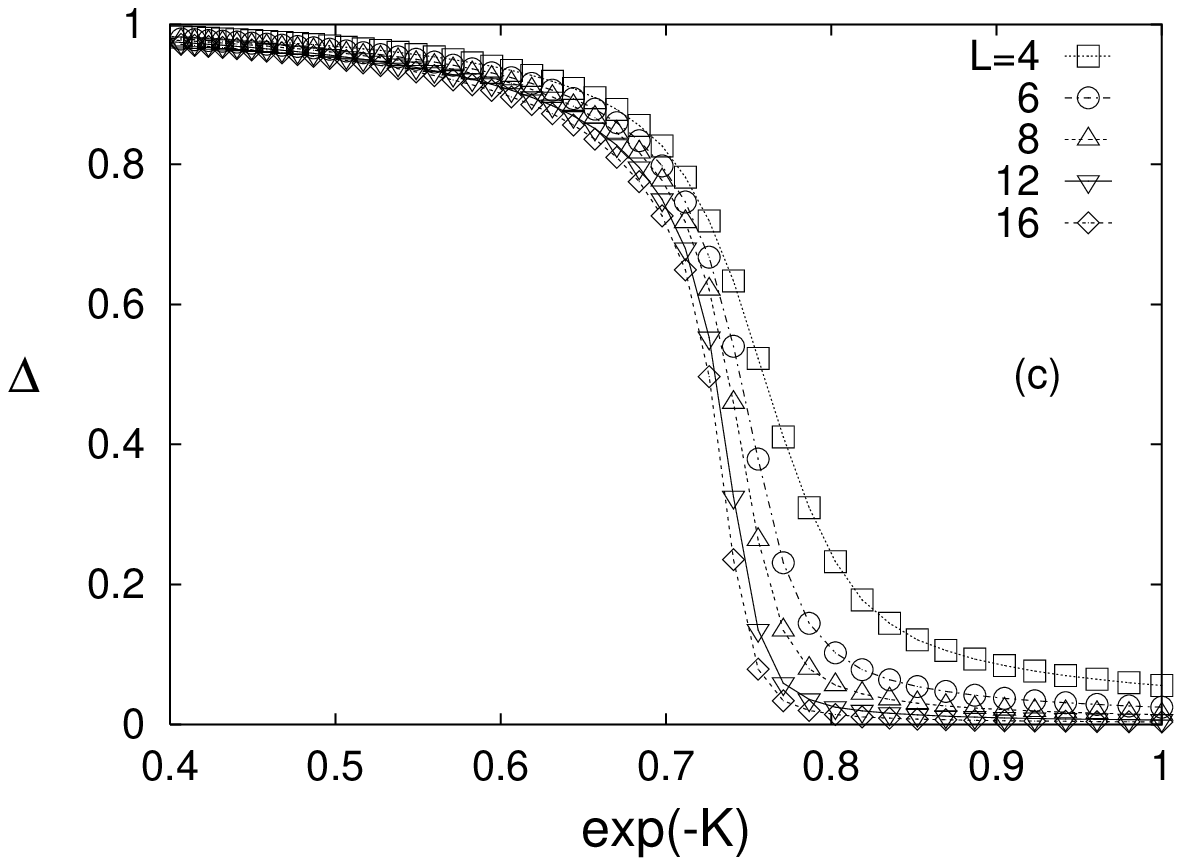}}}
\centering{\resizebox*{!}{6.0cm}{\includegraphics{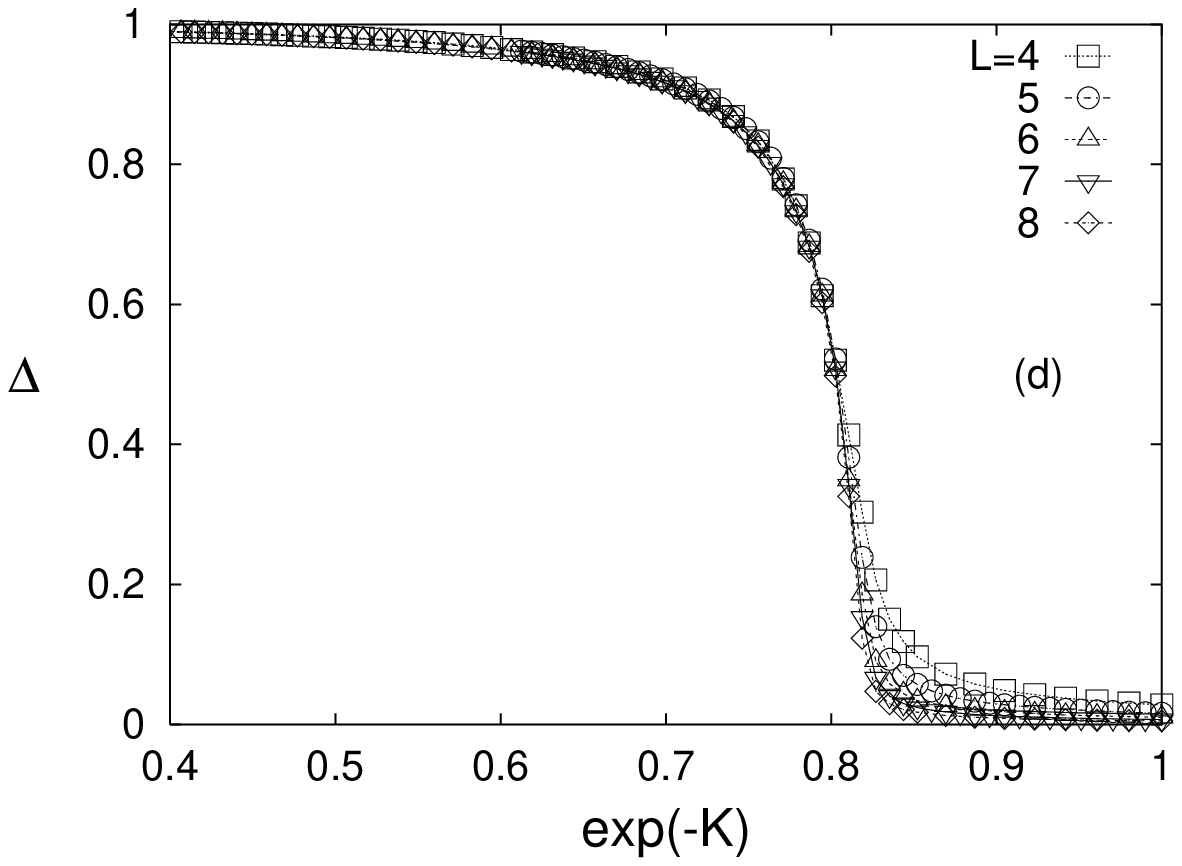}}}
\caption{Phase order parameter $\Delta$ plotted as a function of ${\rm{exp}}(-K)$, where 
$K$ is the coupling strength, with the linear size $L$ varied for $d = $ (a)\,2, (b)\,3, (c)\,4,
and (d)\,5.
}
\label{fig:ph_2345D}
\end{figure}
\begin{figure}
\centering{\resizebox*{!}{6.0cm}{\includegraphics{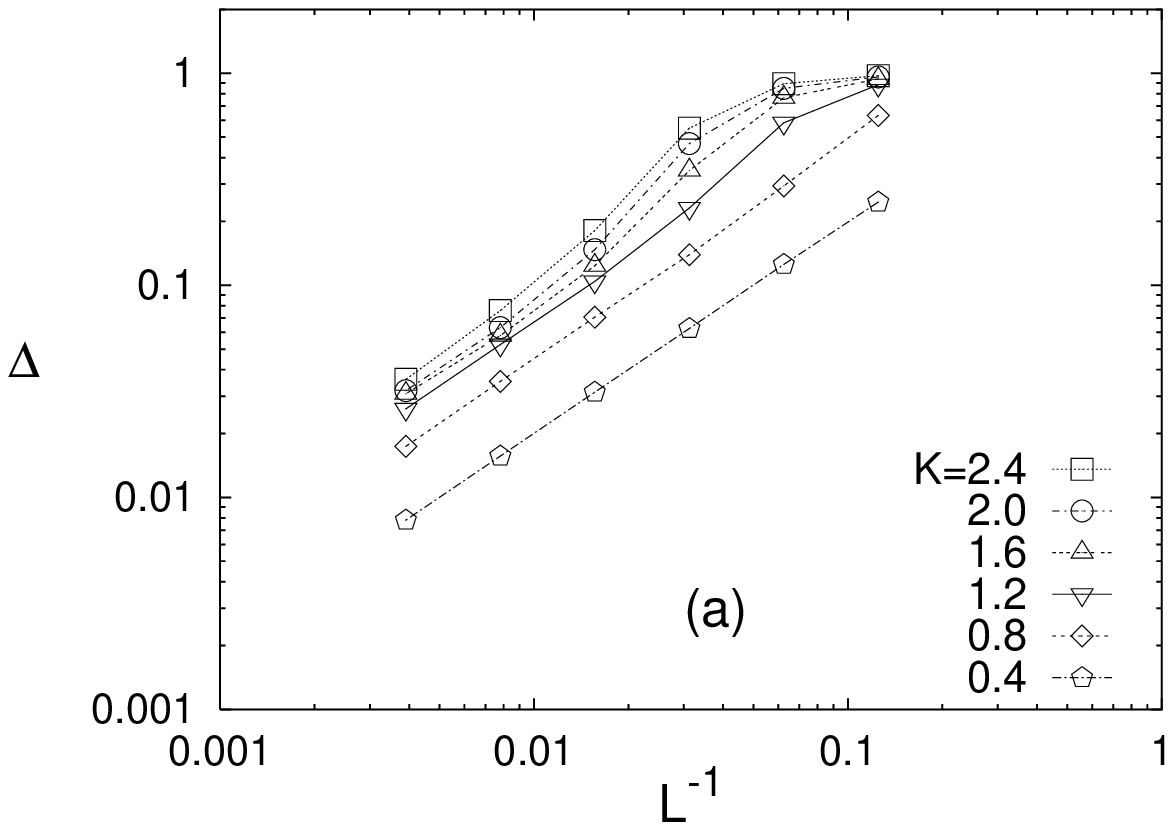}}}
\centering{\resizebox*{!}{6.0cm}{\includegraphics{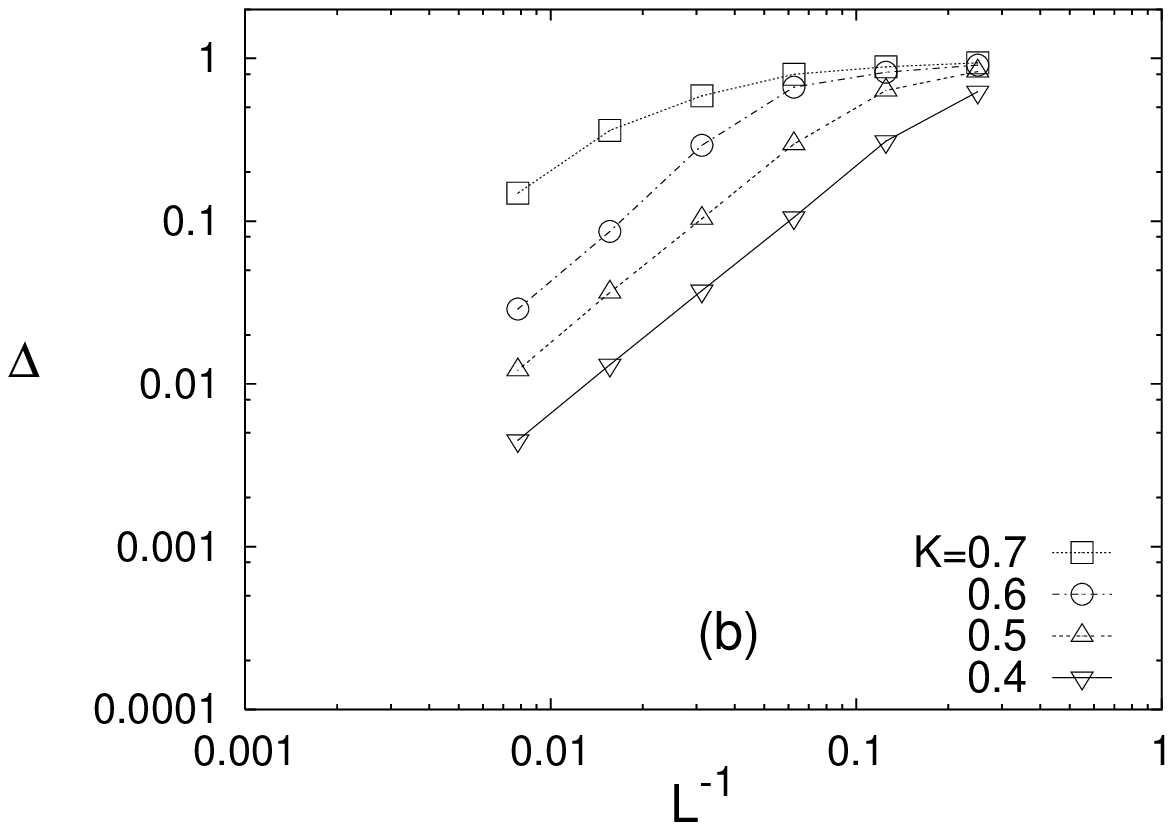}}}
\centering{\resizebox*{!}{6.0cm}{\includegraphics{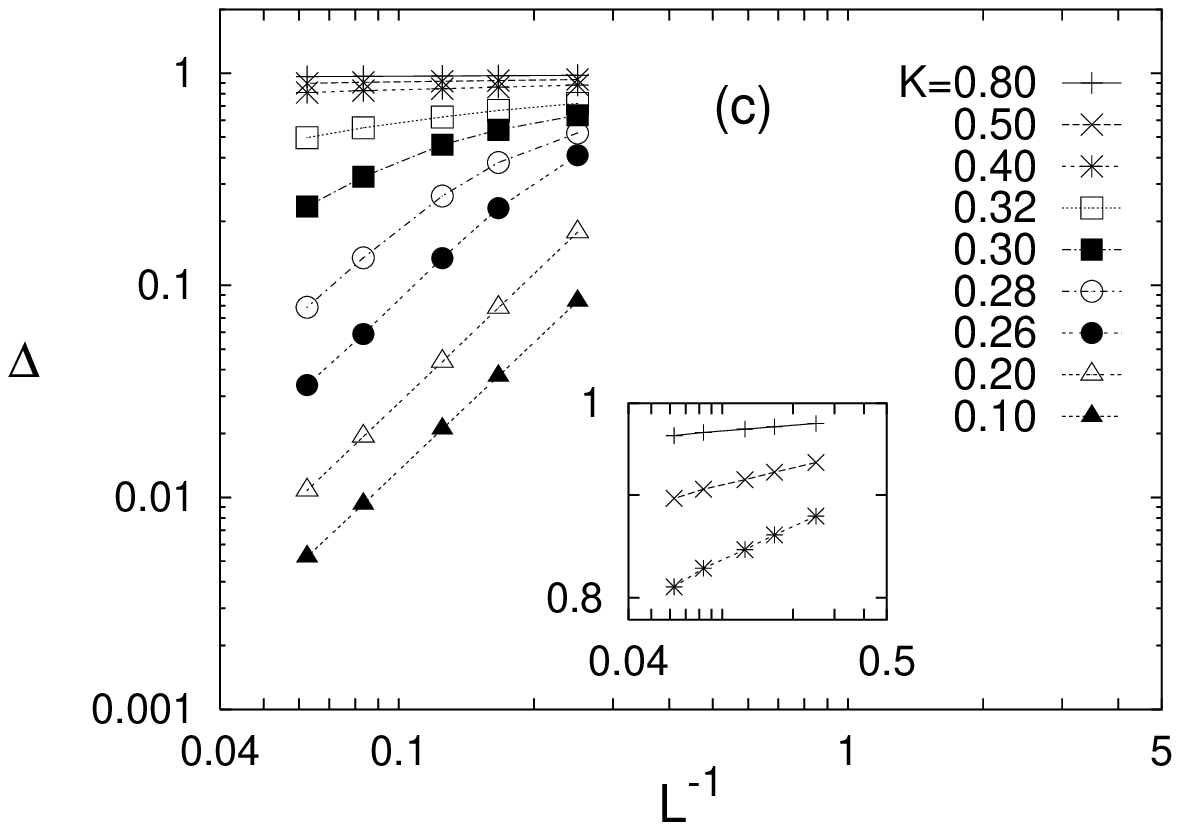}}}
\centering{\resizebox*{!}{6.0cm}{\includegraphics{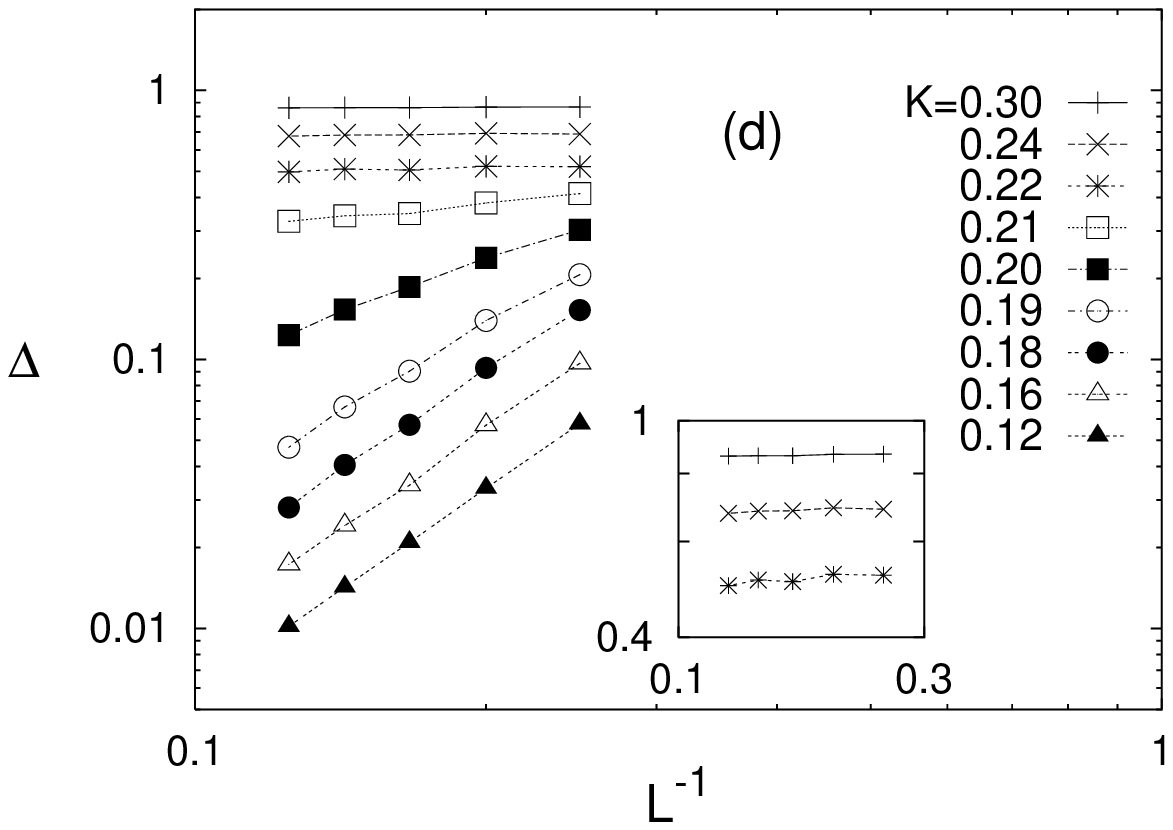}}}
\caption{Log-log plot of the phase order parameter $\Delta$ versus the inverse size $L^{-1}$ 
at various values of the coupling strength $K$ 
for $d = $ (a)\,2, (b)\,3, (c)\,4, and (d)\,5.  Detailed behaviors are shown in the insets.}
\label{fig:ph_L_2345D}
\end{figure}

Figure~\ref{fig:ph_2345D} displays the behavior of the phase order 
parameter $\Delta$ as a function of $\exp (-K)$ for $d=2$ to 5.  For $d=2$ and 3 
[Figs.~\ref{fig:ph_2345D}(a) and (b)], it is evident that the phase order 
parameter $\Delta$ decreases rapidly as the system size $L$ is increased. 
It appears to approach zero in the thermodynamic limit for any finite $K$. 
On the other hand, the size dependence of the phase order parameter 
for $d=4$ and $5$ [see Figs.~\ref{fig:ph_2345D}(c) and (d)] 
is very different from that for $d=2$ or $3$. One is thus tempted to conclude that 
$\Delta$ may approach a nonzero value in the thermodynamic limit 
for large $K$, where the synchronized phase emerges. 

We analyze our data in detail by means of finite-size scaling and show in 
Fig.~\ref{fig:ph_L_2345D} the log-log plots of $\Delta$ versus $L^{-1}$ 
at various values of $K$. 
For $d=2$ in Fig.~\ref{fig:ph_L_2345D}(a), we observe that $\Delta\sim L^{-1}$ 
as expected, which is a characteristic of the fully random phase dominated by runaway
oscillators, up to very large values of $K$ [e.g., $e^{-K}\approx 0.03$]. 
We believe that this fully random phase should extend to arbitrarily large values of $K$, 
as suggested in previous subsections. 

For $d=3$ in Fig.~\ref{fig:ph_L_2345D}(b), this fully random phase seems to terminate
at a finite value of $K$. Our data are consistent with 
the prediction for the fully random phase, $\Delta\sim L^{-3/2}$, in the weak-coupling
regime ($K<K_0$) and with the prediction for the correlated random phase,
$\Delta\sim \exp [-(\sigma/4\pi^3 K^2) L]$, in the strong-coupling regime ($K>K_0$) 
where the linear theory applies. 
The transition point between the fully random phase to the correlated random one
may be determined by the stability analysis of the linear theory. 
The average nearest-neighbor phase difference $G(K)$ defined by Eq.~(\ref{eq:G2}) can be 
used to determine the coupling strength $K_0$ at the transition point
according to $G(K_0)\approx {\cal {O}}(1)$.   Numerically, we find that 
$K_0\approx \sqrt{2\sigma/\pi}$, which is equivalent to the analytical estimate 
by setting $G^{2}(K_0)=1/2$. 
However, in order to firmly establish the correlated random phase in the thermodynamic limit
and explore nature of the transition into the fully random phase, one needs much more extensive
numerical simulations as well as possibly perturbation theory (or stability analysis 
of the fully random phase) in the weak-coupling limit, which are beyond the scope of this paper. 

For $d=4$, our data in Fig.~\ref{fig:ph_2345D}(c) seem to suggest that 
$\Delta$ remains finite for large $K$, even in the thermodynamic limit. 
However, our analytic argument based on the linear theory excludes the
possibility of a nonzero value of $\Delta$ at any finite value of $K$ 
in the thermodynamic limit. In order to resolve this apparent puzzle,
we analyze our data carefully by means of finite-size scaling in Fig.~\ref{fig:ph_L_2345D}(c). 
Manifested for $K\lesssim 0.28$ is the fully random phase: $\Delta\sim L^{-2}$. 
For $K\gtrsim 0.40$, $\Delta$ still decreases algebraically with $L$, as shown in
the inset of Fig.~\ref{fig:ph_L_2345D}(c): $\Delta\sim L^{-\delta(K)}$. 
It is pleasing that our data for $K\gtrsim 0.40$ agree perfectly with
the prediction of the linear theory, $\delta(K)=\sigma/8\pi^2K^2$
(see the previous subsection).
This result confirms that there is no synchronized phase at any finite $K$ 
for $d=4$.  It would also be interesting to explore the possibility of a phase 
transition near $K\approx K_0= \sqrt{\sigma/4}\approx 0.35$ between the fully 
random phase and the critical phase described by the linear theory; this is currently 
under investigation. 

For $d=5$, it looks evident in Fig.~\ref{fig:ph_2345D}(d) that there exists an 
ordered (synchronized) phase extended to finite values of $K$. 
The log-log plot of $\Delta$ versus $L^{-1}$ in Fig.~\ref{fig:ph_L_2345D}(d)
confirms clearly the existence of the synchronization phase transition. 
For $K\lesssim 0.19$, we find the fully random phase: $\Delta\sim L^{-5/2}$. 
For $K\gtrsim 0.21$, on the other hand, the inset of Fig.~\ref{fig:ph_L_2345D}(d)
demonstrates that $\Delta$, first decreasing slightly with $L$,
eventually converges to a non-zero value. 
In fact, for $K\gtrsim 0.24$, this saturated value coincides perfectly 
with the linear-theory value: $\Delta=\exp[-\sigma/(12\pi^2K^2)]$. 
Note here that the linear theory breaks down for
$K\lesssim K_0=\sqrt{\sigma/9}\approx 0.24$ and
the transition into the fully random phase apparently occurs a little later
at $K_c\approx 0.20$. 

The stable ordered (synchronized) phase begins to emerge at $d=5$, while the case 
$d=4$ is marginal, apparently displaying the critical phase; this suggests 
that the lower critical dimension for the phase synchronization is $d_l^P=4$. 
We also note that no complete phase synchronization $(\Delta=1)$ is observed
at any finite $K$ at least up to $d=6$ (see Fig.~\ref{fig:ph_L_2345D}).

\subsection{Phase synchronization transitions for 5$d$ and 6$d$} 

In this subsection we investigate the nature of the phase synchronization 
transition in five ($d=5$) and six dimensions ($d=6$). 
We numerically estimate the values of the critical exponents 
which characterize the universality class of the phase transition.  
In particular, much attention is paid to obtaining 
the critical exponents $\beta$ and $\nu$ which describe the
critical behavior of the order parameter and the correlation 
length, respectively:
\begin{equation}
\Delta \sim (K-K_c)^{\beta} \quad {\rm and}\quad \xi \sim |K-K_c|^{-\nu}, 
\label{eq:crit_behavior}
\end{equation}
where $K_c$ is the critical coupling strength at the transition.

In a finite system, we assume the finite-size scaling relation 
\begin{equation}
\Delta = L^{-\beta/\nu}f[(K-K_c)L^{1/\nu}],
\label{eq:FSS}
\end{equation}
where the scaling function behaves $f(x)\sim x^{\beta}$ as 
$x\rightarrow +\infty$ and $f(x)\sim \mbox{constant}$ as $x\rightarrow 0$.
At criticality $(K=K_c)$, it reduces to 
\begin{equation}
\Delta(K_c,L)\sim L^{-\beta/\nu}.
\label{eq:betanu}
\end{equation}
To estimate efficiently the exponent $\beta/\nu$ and the transition 
point $K_c$, we introduce the size-dependent effective exponent
\begin{equation}
\frac{\beta}{\nu (L)} = - \frac{\ln [\Delta(L^\prime)/\Delta(L)]}{\ln(L^\prime/L)},
\end{equation}
where we take $L^\prime=L+1$ here.  In the thermodynamic limit the value of the 
effective exponent is expected to approach zero for the ordered phase ($K>K_c$), 
$\beta/\nu$ at the transition ($K=K_c$), and $d/2$ for the fully random phase ($K<K_c$), 
respectively.

\begin{figure}
\centering{\resizebox*{!}{5.7cm}{\includegraphics{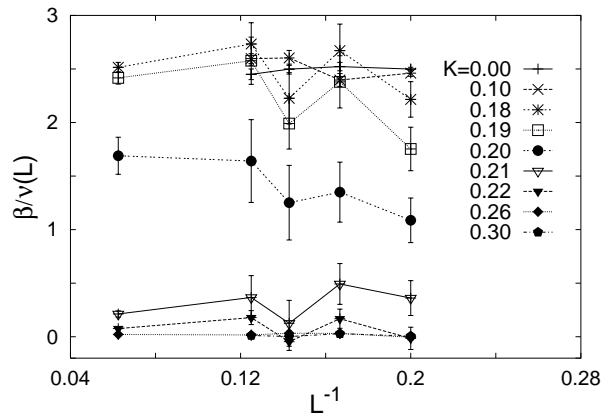}}}
\caption{Effective exponent $\beta/\nu(L)$ versus $L^{-1}$ for $d=5$
at various values of $K$.
}
\label{fig:5D_beta_nu}
\end{figure}

In Fig.~\ref{fig:5D_beta_nu}, we plot the effective exponents computed at
various values of $K$ versus $L^{-1}$ for $d=5$. 
The data for $K\lesssim 0.19$ are observed to converge to the weak-coupling value $5/2$, 
while those for $K\gtrsim 0.21$ converge to zero within statistical errors. 
Only the data at $K=0.20$ appear to converge to a nontrivial value. 
We thus estimate the critical coupling strength $K_c=0.200(5)$ and
the exponent ratio $\beta/\nu=1.6(3)$. 

%
\begin{figure}
\centering{\resizebox*{!}{6.0cm}{\includegraphics{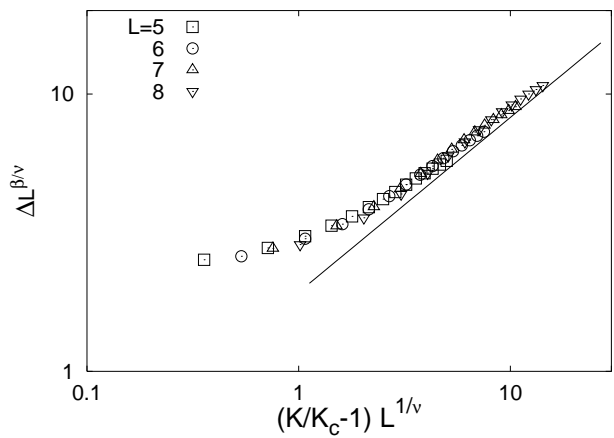}}}
\caption{Data collapse of $\Delta L^{\beta/\nu}$ against $(K/K_c -1)L^{1/\nu}$
in the log-log scale for various values of the system size and coupling strength. 
The best collapse is achieved with $\beta/\nu=1.4(3)$ and $\nu=0.45(10)$.
The straight line has the slope 0.63, giving an estimation of $\beta$.}
\label{fig:5D_scaling}
\end{figure}

To check the finite-size scaling relation directly, we plot
$\Delta L^{\beta/\nu}$ versus $(K/K_c -1)L^{1/\nu}$ in Fig.~\ref{fig:5D_scaling}
and find that the data for various values of $L$ and $K$ are collapsed best
to the curve with the choice $K_c=0.200(5)$, $\beta/\nu=1.4(3)$ and $\nu=0.45(10)$,
which results in $\beta=0.63(20)$.  As expected, the resulting scaling function $f(x)$
converges to a constant for small $x$, and diverges as $x^\beta$ for large $x$
(see Fig. \ref{fig:5D_scaling}). 
Our results for $d=5$ are thus summarized:
\begin{equation}
\beta/\nu =1.5(3),~~ \nu=0.45(10),~~ K_c =0.200(5). 
\end{equation}
We note that there apparently exist substantial deviations from the mean-field (MF) values,
$\beta/\nu=1$ and $\nu=1/2$, although the latter may not be totally excluded. 
In view of the argument for the MF nature,
these apparent deviations are rather unexpected and their origin is unclear at this stage.

\begin{figure}
\centering{\resizebox*{!}{5.7cm}{\includegraphics{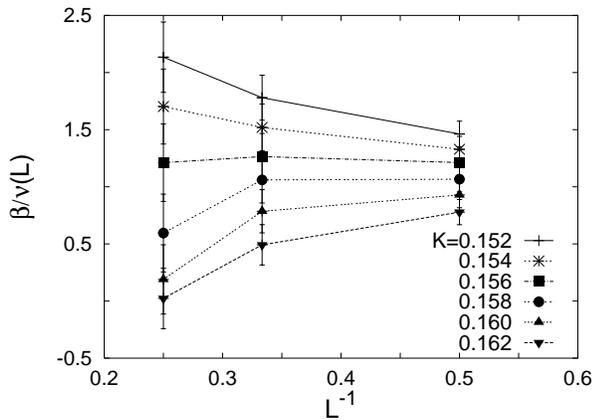}}}
\caption{Effective exponent $\beta/\nu(L)$ versus $L^{-1}$ for $d=6$
at various values of $K$.}
\label{fig:6D_betanu}
\end{figure}

Similarly, Fig.~\ref{fig:6D_betanu} displays the plot of the effective exponent 
$\beta/\nu (L)$ versus $L^{-1}$ for $d=6$, which leads to the estimation:
\begin{equation}
\beta/\nu =1.0(3),~~ \nu=0.45(10),~~ K_c =0.156(2).
\end{equation}
These exponent values appear consistent with the MF values, but 
due to rather large statistical errors it may not be conclusive that 
the $d=6$ system exhibits MF-type critical behavior. 
Further investigations are requisite for determining unambiguously
the upper critical dimension for phase synchronization. 

\section{Frequency entrainment}

In this section, we explore the frequency entrainment, which is 
another kind of synchronization behavior appearing in coupled oscillator systems.
At zero coupling ($K=0$), the phase of each oscillator evolves with its intrinsic 
frequency $\omega_i$: $d\phi_i / dt = \omega_i$. Hence, the frequency (phase velocity) 
distribution of the oscillators remains unchanged from the initial random distribution 
$g(\omega)$. 
With finite local ferromagnetic coupling ($K>0$), the oscillators may tend to 
form locally ordered regions where they evolve with an identical frequency. 
The competition between the ferromagnetic coupling and the dispersion of intrinsic
frequencies sets the size of the locally ordered regions. For sufficiently strong coupling,
the locally ordered regions may expand and merge together into a globally ordered region and 
the number of oscillators with an identical frequency become macroscopic. In this case,
frequencies are said to be entrained macroscopically. 

In characterizing the collective behavior of oscillators in frequency, 
two different frequency order parameters are introduced.
In most previous (numerical) 
studies~\cite{ref:Sakaguchi,ref:Daido,ref:Bahiana,ref:Aoyagi,ref:Strogatz},
a frequency order parameter has been defined to be
\begin{equation}
r \equiv \lim_{N\rightarrow\infty} \left\langle N_s/N \right\rangle,
\label{eq:FOr}
\end{equation}
where $N_s$ and $N$ are the number of oscillators in the largest cluster 
having an identical frequency and the total number of oscillators, respectively.
Nonzero values of $r$ implies the emergence of a macroscopic cluster of oscillators 
with an identical frequency. 
With this definition, the frequency-entrainment transition may be viewed as 
a percolation-type phase transition with a continuous fluctuating variable (frequency). 
In order to measure $r$, we need geometric information on how the oscillators 
with an identical frequency are placed on the $d$-dimensional hypercubic lattice.

In the globally connected oscillator system, there is generically no geometric 
information on the placement of oscillators, each being connected to every other 
oscillator. Therefore, in this case, $N_s$ represents simply the maximum number of oscillators 
with an identical frequency. For later use, we define another frequency order parameter 
without geometrical information as
\begin{equation}
Q \equiv \lim_{N\rightarrow\infty} \left\langle N_n/N \right\rangle,
\label{eq:FOQ}
\end{equation}
where $N_n$ is the maximum number of oscillators with an identical frequency. 
In the system of globally connected oscillators, both definitions are equivalent: 
$r=Q$. Near the frequency-entrainment 
transition, it is known that $Q\sim  (K-K_c)^{\beta}$ with $\beta=1/2$ and 
$K_c=2/\pi g(0)$~\cite{ref:Kuramoto}. Notice that  for globally coupled oscillators
both the phase synchronization and the frequency entrainment transition occur
at the same coupling strength  and their order parameter exponents share the same value.

For the locally connected oscillators, on the other hand, these two 
definitions are not identical. 
The order parameter $Q$ is always larger than $r$, because the former counts 
additional oscillators with an identical frequency belonging to different clusters.
We presume that the order parameter $Q$ should be more suitable for describing 
the frequency-entrainment transition as an order-disorder transition. 
In general, the percolation-type transition characterized by $r$ may or may not occur 
simultaneously at the same coupling strength with the
order-disorder-type transition characterized by $Q$. 
However, in our model with a continuous fluctuating variable, it may be reasonable to assume 
that these two order parameters behave similarly, at least qualitatively near 
the transition from the ordered side. 

Here we measure the order parameter $Q$ to probe the frequency-entrainment transition
for simplicity and convenience. Accordingly, it is not necessary to retain geometrical 
information during integration of Eq.~(\ref{eq:model}). 

\subsection{Linear theory}

Similarly to the case of phase synchronization, we begin with the linearized equation of motion
in Eq.~(\ref{eq:linear}) and measure the fluctuation width of the growing velocity 
(rather than the height) of the surface, which would provide key information on 
the dispersion of the phase velocity (frequency). 
The mean-square fluctuation width for the phase velocity 
$v({\bf{x}},t)\, [\equiv \dot\phi({\bf{x}},t)]$ is defined to be
\begin{equation}
V^2(t) \equiv \frac{1}{L^d}\int^{L} d^d {\bf{x}}
\left\langle \left[ v({\bf x},t)- {\bar v}(t) \right]^2 \right\rangle,
\label{eq:Wv}
\end{equation}
where ${\bar v}(t)$ is the spatial average.

Taking the time derivative of the Fourier-space solution in Eq.~(\ref{eq:solution}), 
we find 
\begin{equation}
v({\bf{k}},t)=\omega({\bf k})\,e^{-Kk^2t}-Kk^2\phi({\bf k},0)\, e^{-Kk^2t}.
\label{eq:dotsolution}
\end{equation}
One can easily see that, at any finite $K$, all Fourier components of the phase velocity 
except the ${\bf k}=0$ mode vanish in the long-time limit ($t\rightarrow\infty$),
which indicates that the phase velocity becomes uniform in space
and $V$ approaches zero. Without coupling ($K=0)$, the velocity distribution should be 
identical to the initial frequency distribution $\omega({\bf k})$ so $V^2$ 
is equal to $2\sigma$ at all times. 
From the above equation, it is straightforward to show the frequency width, normalized to 
the fully random value:
\begin{equation}
\frac{V^2}{2\sigma} = \Omega_d \int_{2\pi/L}^{\pi/a} dk k^{d-1}e^{-2Kk^2 t}.
\label{eq:V2}
\end{equation}
In the short-time regime $(K t \ll L^2)$, it decays algebraically for any nonzero $K$ 
in any space dimension $d$,
\begin{equation} 
\label{eq:Wv_result}
\frac{V^2}{2\sigma}  \sim t^{-d/2},
\end{equation}
and eventually vanishes in the regime $K t \gg L^2$, as expected. In the language of 
surface growth, this corresponds to the completely flat phase. 

The uniform distribution of the phase velocity $(V=0)$ implies complete frequency entrainment. 
Therefore, our linear theory predicts that, in the stationary state, 
the frequencies are completely entrained (with the frequency order parameter $r=Q=1$) 
for any nonzero $K$ in any space dimension. 
Only is there a trivial first-order transition at $K=0$ from the fully random phase 
($r=Q=0$) to the completely entrained phase. Of course, any prediction for $d\le 2$ 
should be untrustworthy because there nonlinear effects dominate in the whole range of $K$,
invalidating the linear theory. 
On the other hand, for $d>2$, the linear theory may survive to establish 
the completely entrained phase for large $K$ where the average nearest-neighbor 
phase difference becomes $G(K)\lesssim {\cal {O}}(1)$.

Finally, we note that, unlike the phase synchronization problem, 
there is no explicit and quantitative relation between the frequency fluctuation width $V$ 
and the frequency order parameter $r$ or $Q$. 

\subsection{Nonlinear effects}

The results from the linear theory  can be understood rather in a simple manner. 
Consider the linearized equation of motion in Eq.~(\ref{eq:linear}) without 
higher-order terms. 
Taking the time derivative of this equation, the disorder term $\omega({\bf x})$ 
drop out and yields a simple noise-free equation 
\begin{equation}
\frac{\partial}{\partial t}v({\bf{x}},t) = K\nabla^2 v({\bf{x}},t).
\label{eq:diff}
\end{equation}
This equation can be viewed as the standard diffusion equation governing heat conduction 
with the diffusion constant $K$ and the phase velocity (frequency) field identified 
as the temperature field. 
Since the diffusion constant $K$ is positive, one can easily expect that any local temperature
gradient should disappear, giving rise to a uniform distribution of the temperature field 
in the long-time limit. In terms of the oscillator language, 
the phase velocity of all oscillators become identical, 
leading to a delta-function-like distribution (i.e., completely entrained phase in frequency). 

Now we come back to the original equation in Eq.~(\ref{eq:model}) 
to accommodate nonlinear effects. 
After taking the appropriate continuum limit in space and dropping off higher-order terms 
(but not expanding the sine function to retain the nonlinear effects at least in lower orders), 
we get
\begin{equation}
\frac{\partial}{\partial t} \phi({\bf{x}},t)= \omega({\bf{x}}) + 
2K\sum_\mu \sin\left(\frac{1}{2}\left({\bf\hat e}_\mu\cdot  {\bf \nabla}\right)^2 \phi\right),
\label{eq:sine_linear}
\end{equation}
where ${\bf\hat e}_\mu$ denotes the unit lattice vector in the $\mu$ direction 
and the summation is over all $d$ different directions. 
The linearized equation in Eq.~(\ref{eq:linear}) is restored by expanding 
the sine function and keeping the lowest order. 
With the rotational symmetry (if not spontaneously broken), 
this equation can be written approximately in a simpler form 
\begin{equation}
\frac{\partial}{\partial t} \phi({\bf{x}},t)= \omega({\bf{x}}) + 
2K \sin\left(\frac{1}{2}\nabla^2 \phi\right),
\label{eq:sine_linear_simple}
\end{equation}
which, upon taking the time derivative, gives 
\begin{equation}
\frac{\partial}{\partial t} v({\bf{x}},t)= 
K \cos\left(\frac{1}{2}\nabla^2 \phi\right)\nabla^2 v.
\label{eq:v_sine_linear_simple}
\end{equation}
This equation can also be viewed as the diffusion (heat conduction)equation, 
with the {\em effective} diffusion constant $K\cos(\frac{1}{2}\nabla^2 \phi)$ varying
in space and time. More importantly, it may become negative depending on 
the value of $\nabla^2 \phi$. With negative diffusion constant, a local thermal
gradient does not diminish but increase and the system can become highly inhomogeneous. 
As the effective diffusion constant changes its sign frequently in time and also in space,
the system may reach, through competition between diffusion and localization 
(negative diffusion), a stationary state with a nonuniform temperature distribution. 
In the oscillator language, the phase velocity can take a broad distribution 
in addition to or in the absence of delta-function-like peaks. 

\begin{figure}
\centering{\resizebox*{!}{5.7cm}{\includegraphics{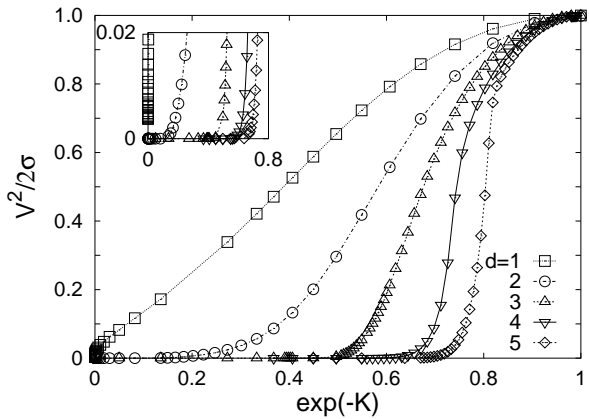}}}
\caption{The mean-square width $V^2$ of the phase velocity, which is divided by 
the variance $2\sigma \,(=1)$, is displayed as a function of ${\rm exp}(-K)$ 
in various dimensional systems.  The detailed behavior near $V^2 =0$ is shown in the inset.}
\label{fig:V2_K}
\end{figure}

The frequency fluctuation width $V$ representing the broadness of the frequency distribution 
is expected to become nonzero in the strong-coupling regime where nonlinear effects 
may become dominant. We integrate the full nonlinear discrete equation 
in Eq.~(\ref{eq:model}) numerically and directly measure $V^2$. 
Figure~\ref{fig:V2_K} displays the normalized mean-square width of the phase velocity 
$V^2/2\sigma$ versus $\exp(-K)$ with $2\sigma=1$ and $L=25600, 128, 64, 16$, and $8$ for
$d=1, 2, 3, 4$, and $5$, respectively.  
At zero coupling, $V^2/2\sigma$ should become unity, while in the completely entrained 
phase it should vanish. We do not find any appreciable finite-size effects
in the whole range of $K$ in all dimensions except for very small values of $V^2/2\sigma$, 
which will be discussed later. 

Observed are various interesting features: 
First, we find that the fully random phase exists only at zero coupling in all dimensions. 
The normalized width $V^2/2\sigma$ becomes smaller than unity as soon as the coupling
is turned on. This result is in sharp contrast to the case of phase synchronization, 
where the fully random phase extends over the whole range of $K$ in one and two dimensions
and also forms a sizable region in the weak-coupling regime in higher dimensions. 
Of course, these two results are not contradictory. 
The oscillator phases cannot be correlated without correlated frequencies; 
however, the reverse is not necessarily true. With frequencies already correlated, the phases
can still be fully random if the coupling between oscillators are weak. 
It should not be mistaken that $V^2/2\sigma < 1$ does not guarantee any macroscopic
frequency entrainment (i.e., any nonzero value of the frequency order parameter $Q$). 
In the case of phase synchronization, a similar situation has been found for $d=3$ and $d=4$,
where phases are correlated but the phase order parameter $\Delta$ remains zero. 
Therefore, it seems that the normalized width does not provide proper information on the
frequency entrainment-detrainment transition. 

Second, the normalized width gradually decreases and approaches zero as $K$ is increased
in all dimensions. The one-dimensional case is special in that the normalized width 
appears finite at any finite $K$, indicating no completely entrained phase, 
and approaches the $K=\infty$ point with a nonzero (very large) slope as a function 
of $\exp (-K)$ (see the inset of Fig.~\ref{fig:V2_K}). 
On the other hand, all other cases $d\ge 2$ appear to possess regions of vanishing width 
($V=0)$ in the strong-coupling regime, where $V^2/2\sigma$ decreases exponentially 
with an infinitesimally small slope at finite values of $K$. 
This implies that the completely entrained phase may be present at finite coupling 
strength for $d\ge 2$, where the linear theory applies. 

\begin{figure}
\centering{\resizebox*{!}{6.0cm}{\includegraphics{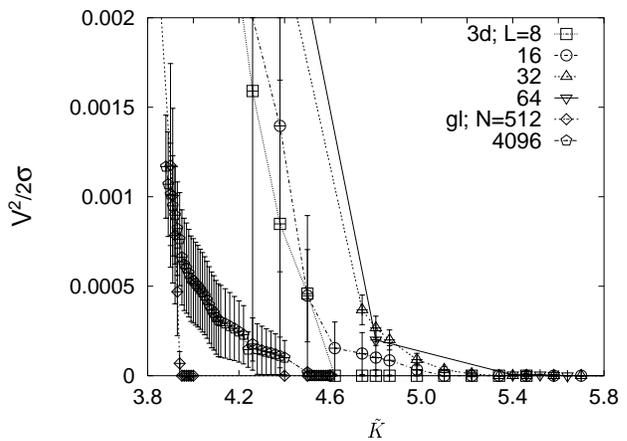}}}
\caption{Behavior of $V^2/2\sigma$ of the phase velocity in various dimensional systems, 
depending on the coupling strength ${\tilde K}$. For comparison, the data for 
the globally coupled system are also shown, represented by {\it gl}.
}
\label{fig:complete_forder}
\end{figure}

However, the stability analysis of  the completely entrained phase should be done 
with great care. The globally coupled system, where each oscillator is coupled with 
every other oscillators with equal strength $K/N$, provides analytic results, 
with which our results may be compared. It is well known that there is no complete 
frequency entrainment at finite $K$ for the globally coupled system, 
given with the intrinsic frequency distribution $g(\omega)$ having no cutoff 
at finite frequency $\omega$ [our choice $g(\omega)\sim \exp (-\omega^2/4\sigma)$
provides an example; see the next section]. 
Figure \ref{fig:complete_forder} displays the behavior of $V^2/2\sigma$ 
versus ${\tilde K}\equiv Kz$ for $d=3$, where $z \,(=2d)$ is the coordination number.
For comparison, the data for the globally coupled system of the same size $(N=L^d)$,
with ${\tilde K}=K$ are also shown. 

As the system size $L$ is increased, the value of ${\tilde K}={\tilde K}_c^L$ beyond which 
the mean-square width for the phase velocity vanishes ($V^2/2\sigma=0$) tends to
shift to larger values, albeit slowly, both in the globally coupled and 
the locally coupled systems for $d=3$. 
The value of ${\tilde K}_c^L$ should diverge to infinity as $L\rightarrow \infty$ 
in the globally coupled system. We do not find much difference between these two systems 
in the value of ${\tilde K}_c^L$ and its finite-size behavior. 
Actually, the value of ${\tilde K}_c^L$ is a little bit larger (corresponding to 
the  narrower completely entrained phase) in the $d=3$ case than in the globally coupled one. 
Based on this observation, we suggest that ${\tilde K}_c^L$ for a locally coupled system 
also grows arbitrarily large in the thermodynamic limit and there may not exist 
complete frequency entrainment at finite $K$ in any dimensions. 
Of course, if the distribution $g(\omega)$ has finite cutoffs 
(like the semi-circle distribution), the locally coupled system might have the 
completely entrained phase at finite values of $K$, similarly to the globally coupled one.

\subsection{Frequency order parameter}

We now measure the frequency order parameter $Q$ defined in Eq.~(\ref{eq:FOQ}),
which is the maximum fraction of the oscillators having an identical frequency. 

Each oscillator starts to evolve with its intrinsic frequency given by the
initial distribution $g(\omega)$, but the coupling between near neighboring
oscillators will modify its frequency continuously with time until the
the stationary state is reached. We measure the mean frequency of the $i$th oscillator
after some transient time $t_s$:
\begin{equation}
\bar\omega_i \equiv \lim_{t\rightarrow\infty} \frac{\phi_i(t)-\phi_i(t_s)}{t-t_s}.
\label{eq:def_modify_w}
\end{equation}

\begin{figure}
\centering{\resizebox*{!}{4.2cm}{\includegraphics{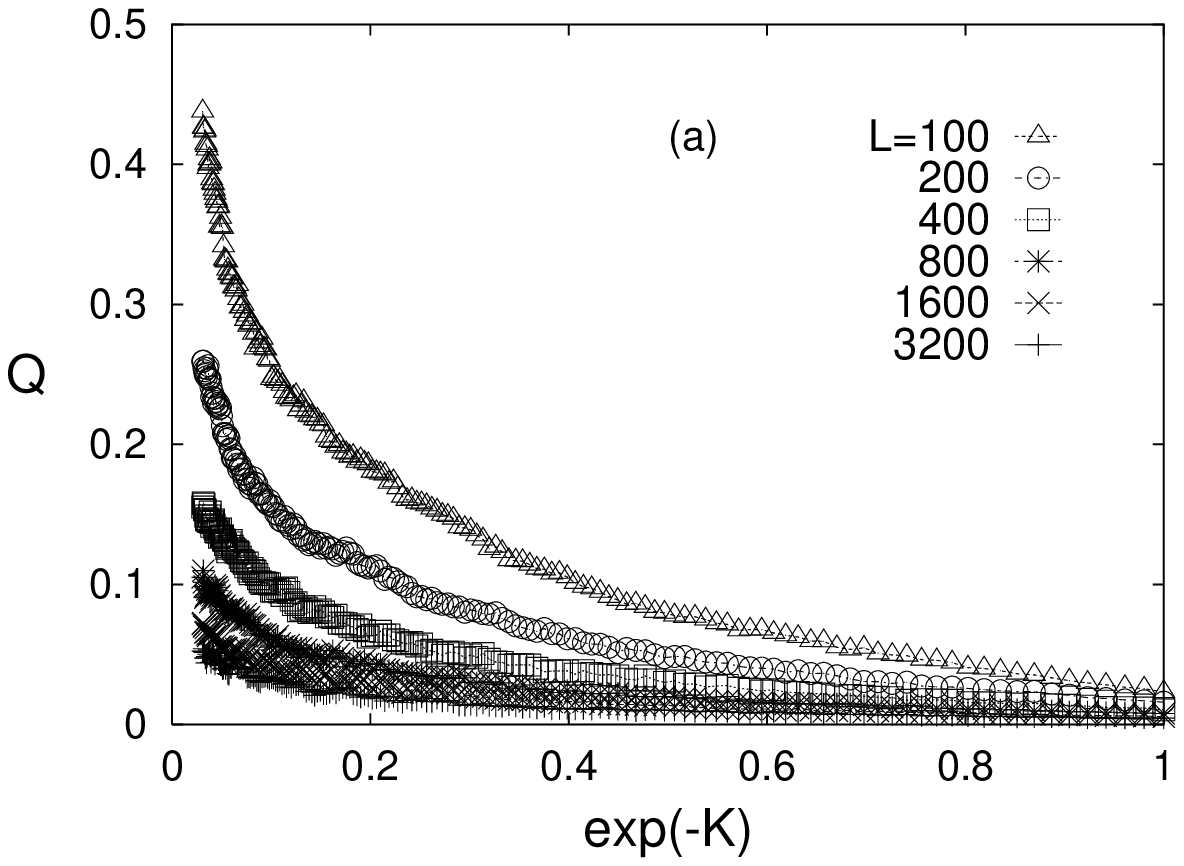}}}
\centering{\resizebox*{!}{4.2cm}{\includegraphics{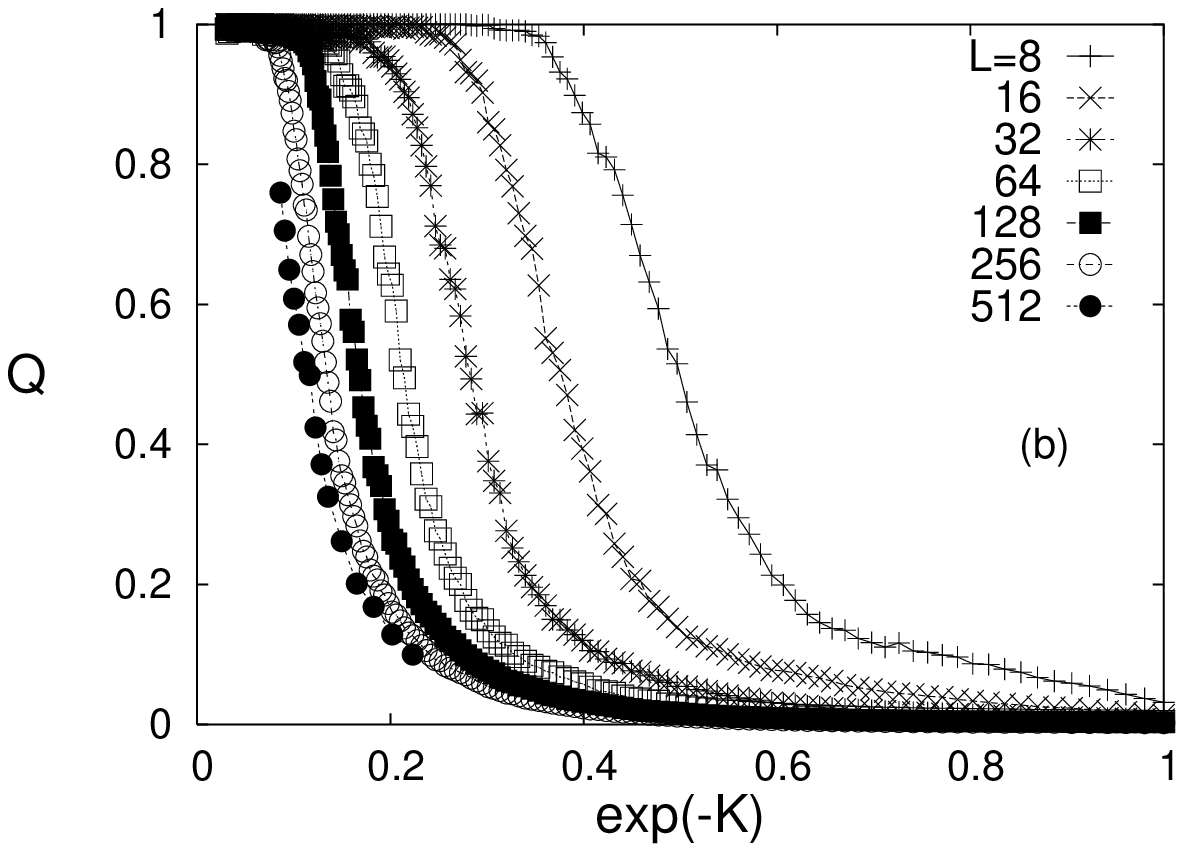}}}
\centering{\resizebox*{!}{4.2cm}{\includegraphics{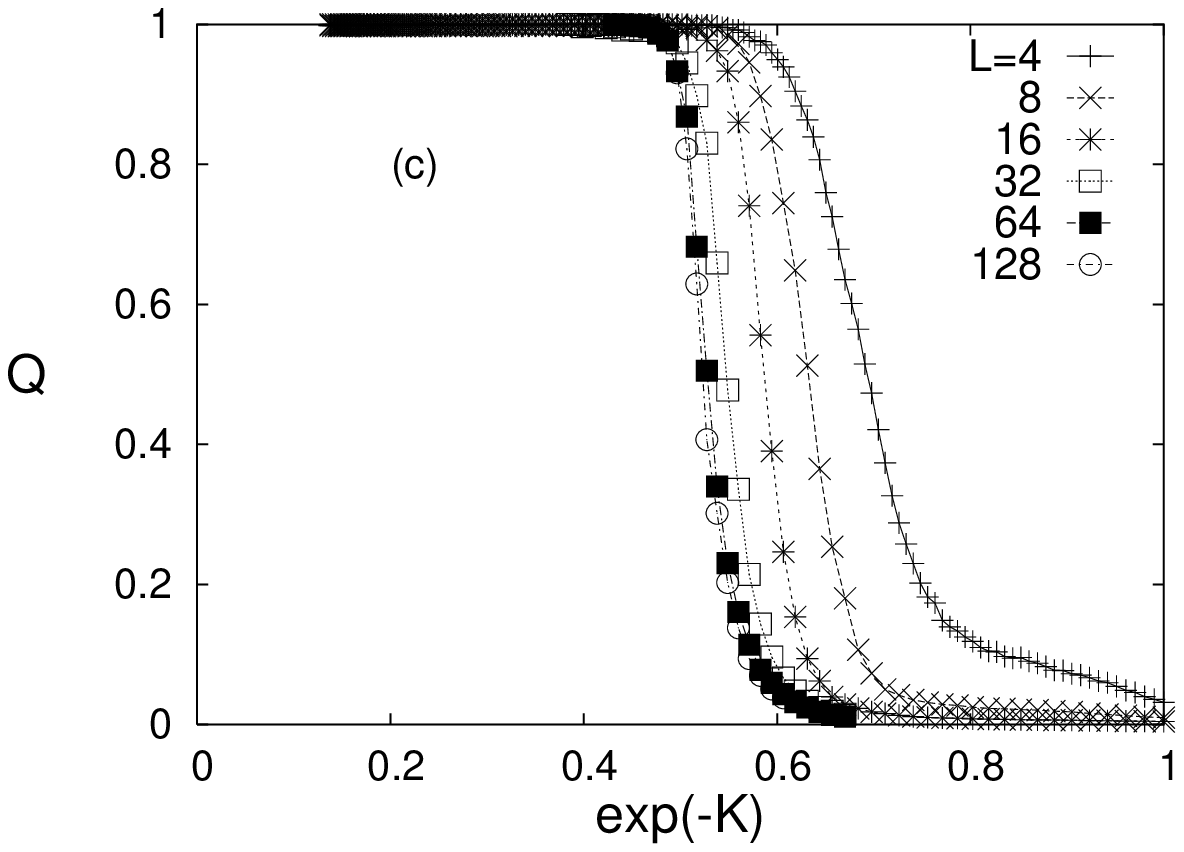}}}
\centering{\resizebox*{!}{4.2cm}{\includegraphics{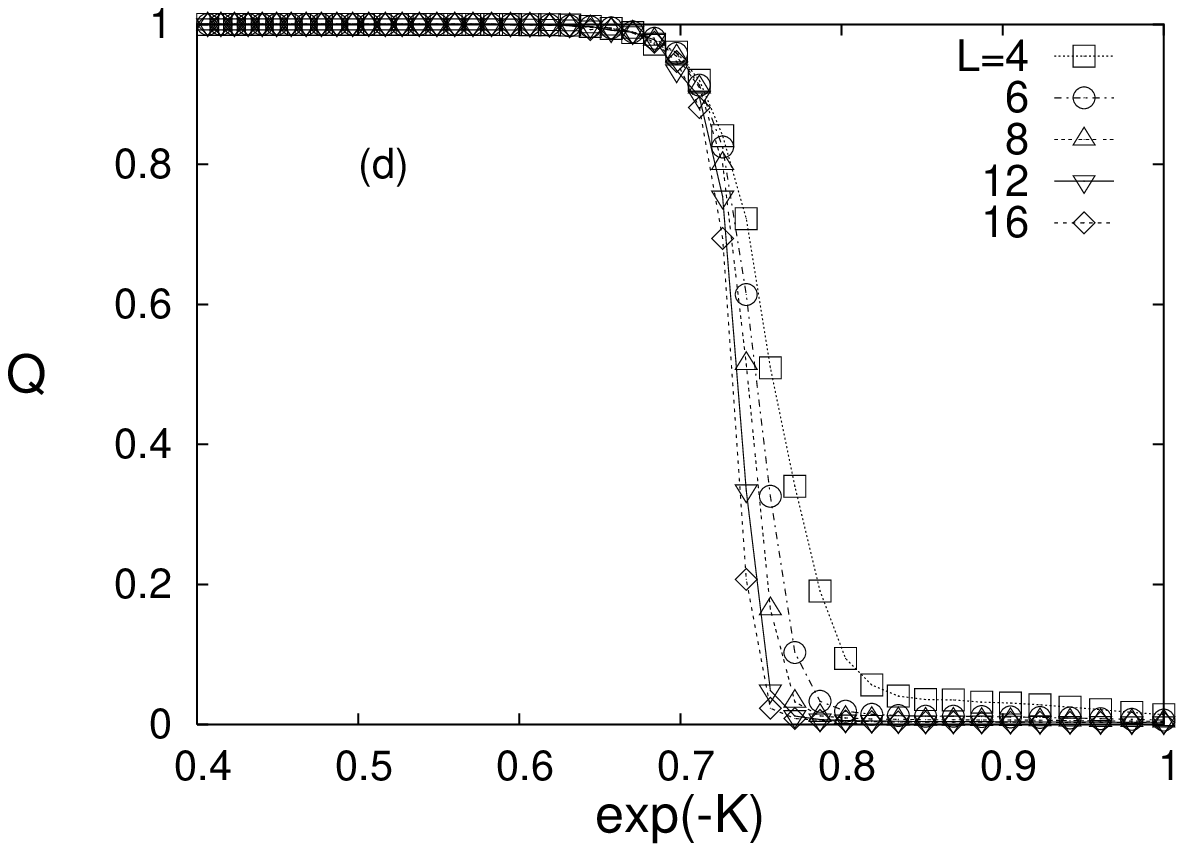}}}
\centering{\resizebox*{!}{4.2cm}{\includegraphics{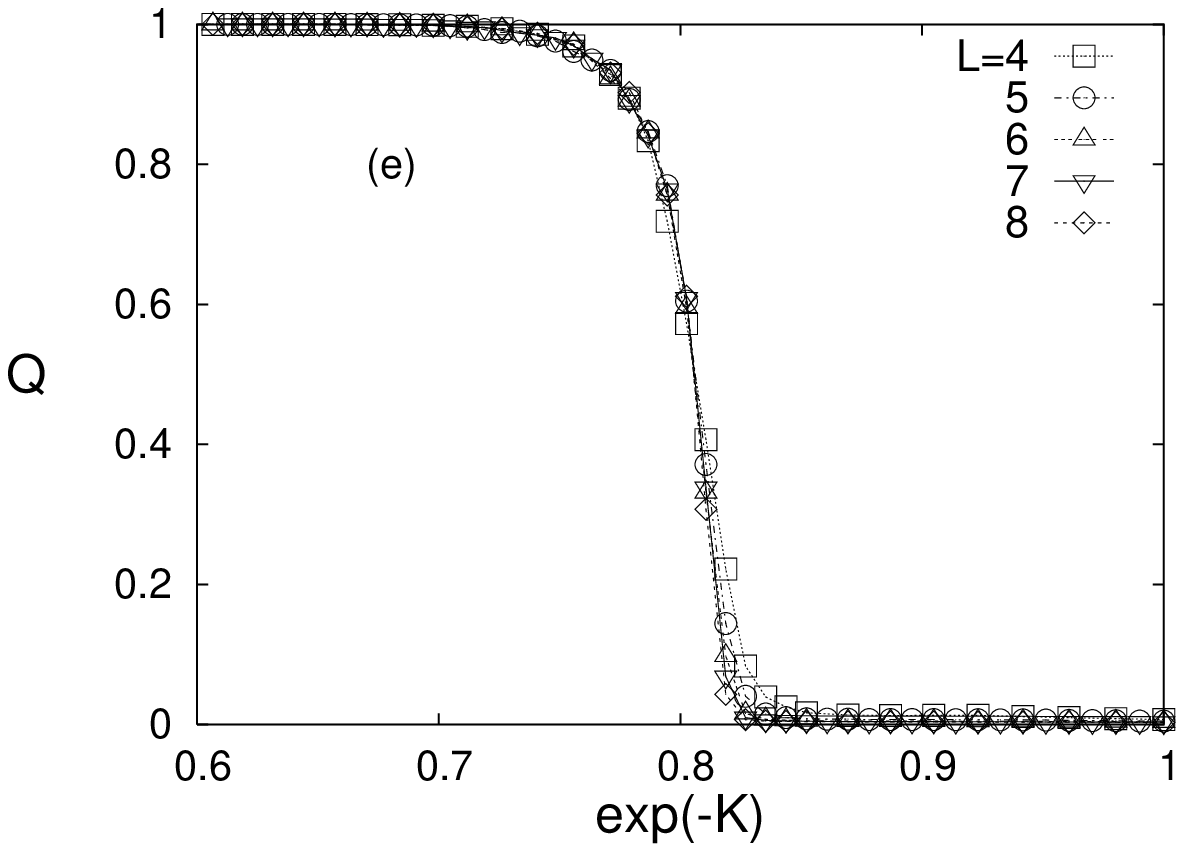}}}
\caption{Frequency order parameter $Q$ plotted as a function of ${\rm exp}(-K)$, 
with the linear size $L$ varied for $d = $ (a)\,1, (b)\,2, (c)\,3, (d)\,4, and (e)\,5.
}
\label{fig:rp_2345D}
\end{figure}

We integrate Eq.~(\ref{eq:model}) up to $N_t=4\times 10^4$ time steps
with time increment $\delta t=0.05$ and measure ${\bar \omega}_i$
at the maximum time $t_m=N_t \delta t$ with the transient time $t_s=2.8\times 10^4 \delta t$.
Since the frequency resolution is limited by  $\delta\omega = \pi (t_m-t_s)^{-1}$ 
in numerical integration, the frequencies of different oscillators are regarded as identical
if the frequency difference does not exceed $\delta\omega \approx 5.2\times 10^{-3}$.
With this resolution, we can draw a histogram $h({\bar \omega})$ of the number of oscillators with
an identical frequency ${\bar \omega}$ normalized by the total number of oscillators, 
i.e.~$\sum_{\bar \omega} h({\bar \omega}) =1$. The order parameter $Q$ is
then obtained from the maximum (peak) value of $h({\bar \omega})$. 
Precisely speaking with finite resolution $\delta\omega$, 
$Q$ is given by the maximum value of $h({\bar \omega})$ minus 
$g(0)\delta\omega=\delta\omega/\sqrt {4\pi\sigma}\approx 2.1\times 10^{-3}$, 
which is the  maximum value at $K=0$. In practice, we measure this quantity, which
should vanish in the detrained phase even with finite resolution. 

Figure \ref{fig:rp_2345D} displays the behavior of the frequency order parameter $Q$ 
as a function of  $\exp (-K)$ for $d=1$ to 5.
For $d=1$ shown in Fig.~\ref{fig:rp_2345D}(a), it is evident that the frequency order 
parameter $Q$ decays rapidly with the system size $L$, seemingly approaching zero 
in the thermodynamic limit for any finite $K$. 
This implies that there is no frequency entrainment at all in one dimension. 
For $d=2$, Fig.~\ref{fig:rp_2345D}(b) shows that $Q$ decreases slowly and 
the entrained phase continues to shrink with $L$. 
For $d=3, 4,$ and 5, it is observed in Figs.~\ref{fig:rp_2345D}(c), (d), and (e) that 
$Q$ appears to saturate to a nonzero value in the strong-coupling regime, 
which suggests that there exists a frequency entrainment-detrainment transition for $d\ge 3$.
Moreover, it appears that the fully entrained phase ($Q=1$) begins to show up 
at finite values of $K$. However, as discussed in the previous subsection, 
the careful analysis in comparison with the globally coupled system suggests that 
complete entrainment occurs only at $K=\infty$. 

\begin{figure}
\centering{\resizebox*{!}{6.0cm}{\includegraphics{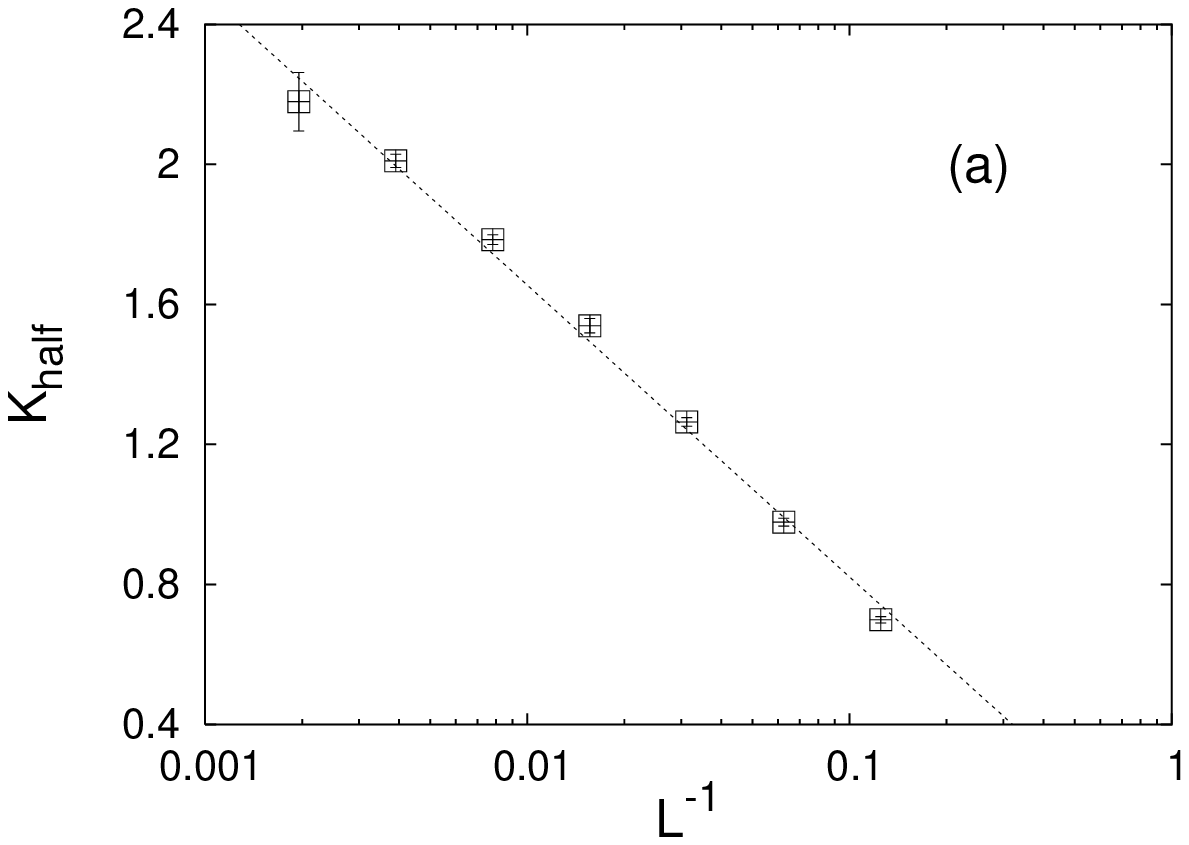}}}
\centering{\resizebox*{!}{6.0cm}{\includegraphics{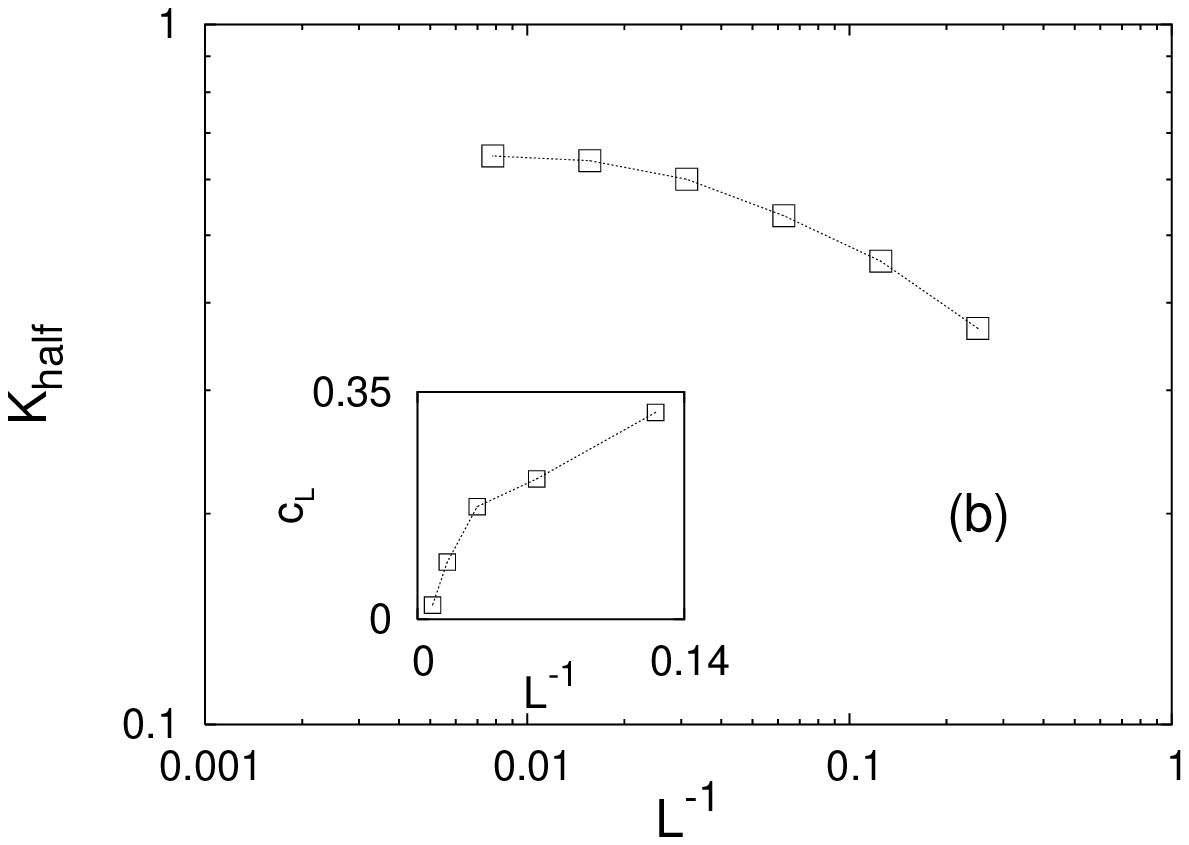}}}
\caption{Half-order coupling strength $K_{\rm half}$ for frequency entrainment in 
(a) two ($d=2$) and (b) three dimensions ($d=3$), plotted against the inverse linear 
size $L^{-1}$ in the semi-log scale and in the log-log scale, respectively. 
For $d=2$,  $K_{\rm{half}}$ diverges logarithmically with $L$: 
$K_{\rm{half}}\simeq a \ln L$ with $a\approx 0.35(15)$. 
For $d=3$, $K_{\rm{half}}$ seems to converge exponentially to a value around $0.73$. 
The inset demonstrates that the successive slope $c_{L}$ between neighboring data 
approaches zero as $L\rightarrow\infty$.
}
\label{fig:q_Keff_23D}
\end{figure}

We analyze our data more carefully to locate the transition point $K_c$ 
separating the entrained phase from the detrained one for $d=2$ and 3. 
We define the half-order value of the coupling strength $K_{\rm half}$ 
at which the frequency order parameter becomes one half: $Q(K_{\rm half})=1/2$ 
and investigate its finite-size behavior. 
In Fig.~\ref{fig:q_Keff_23D}(a), the half-order value $K_{\rm half}$ for $d=2$ is 
plotted versus $L^{-1} $ in the semi-log scale, which displays logarithmic 
divergence $K_{\rm{half}}\simeq a\ln L$ with $a\approx 0.35(15)$. 
This confirms that there is no entrained phase presented in two dimensions. 
It is of interest to note the behavior $K_{\rm{half}}\sim {(\ln L)}^{1/2}$ in the case of 
phase synchronization for $d=4$ (see Sec.~III). 

In three dimensions, the situation is clearly different from that in two dimensions. 
Figure~\ref{fig:q_Keff_23D}(b) displays the log-log plot of $K_{\rm{half}}$ versus $L^{-1}$.
As $L\rightarrow \infty$, $K_{\rm{half}}$ is observed to converge to a finite value 
around $0.73$. The analysis on the successive slope $c_L$ between data points 
for subsequent sizes confirm this convergence. 
Therefore, we conclude that there is a frequency entrainment-detrainment transition 
at $K=K_c\lesssim 0.73$ in three dimensions. 
Similar conclusions can be drawn in higher dimensions as well. 

\begin{figure}
\centering{\resizebox*{!}{4.2cm}{\includegraphics{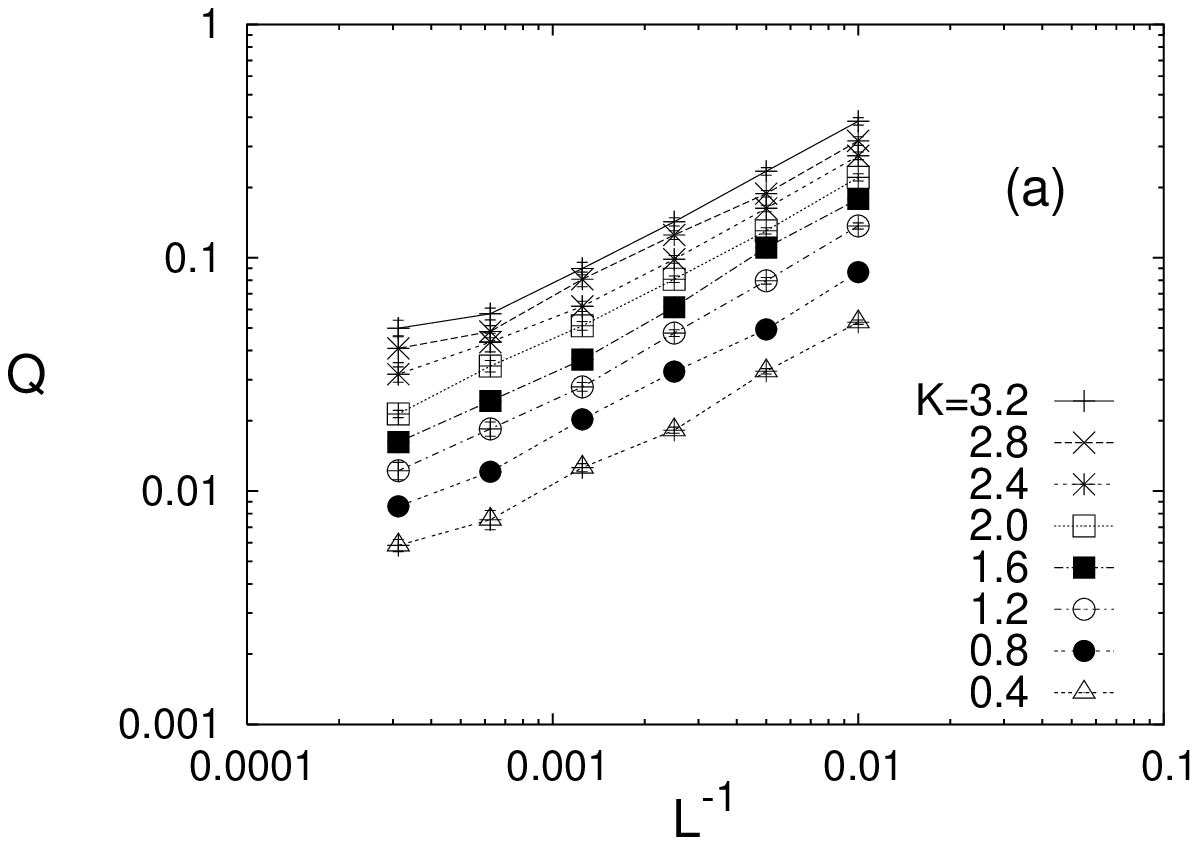}}}
\centering{\resizebox*{!}{4.2cm}{\includegraphics{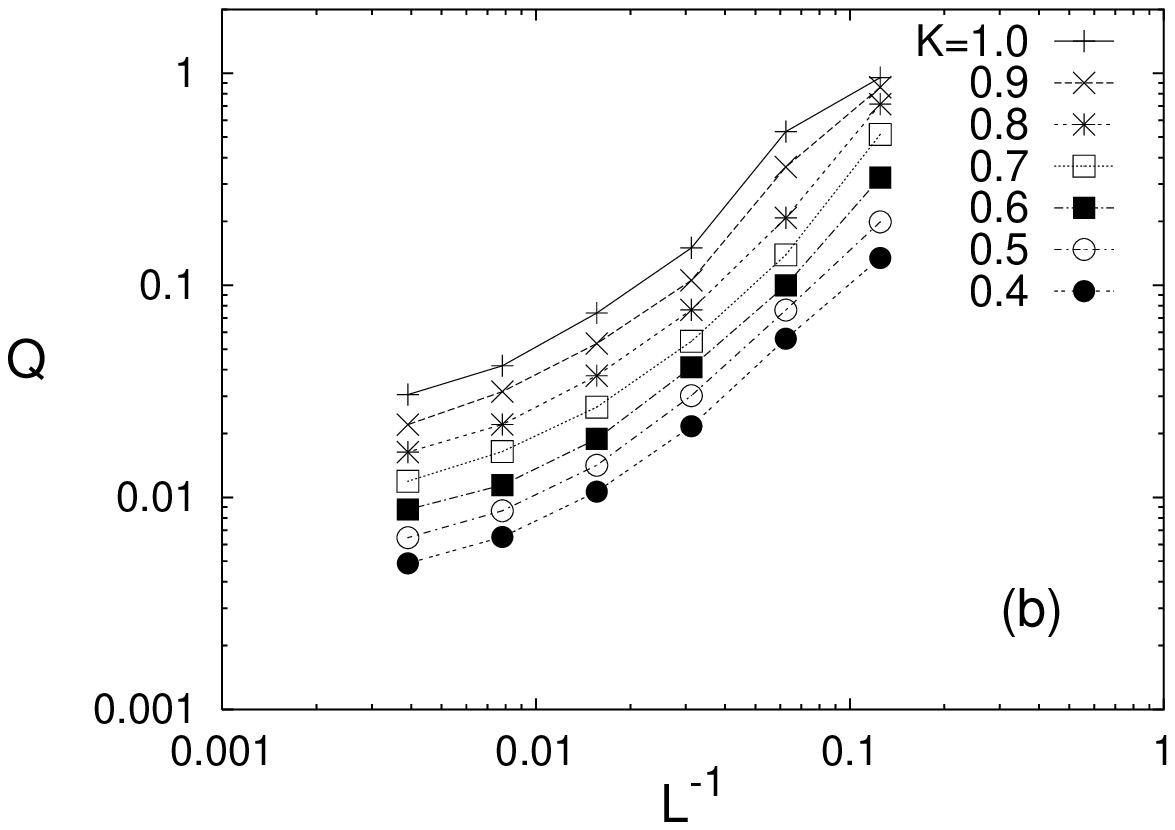}}}
\centering{\resizebox*{!}{4.2cm}{\includegraphics{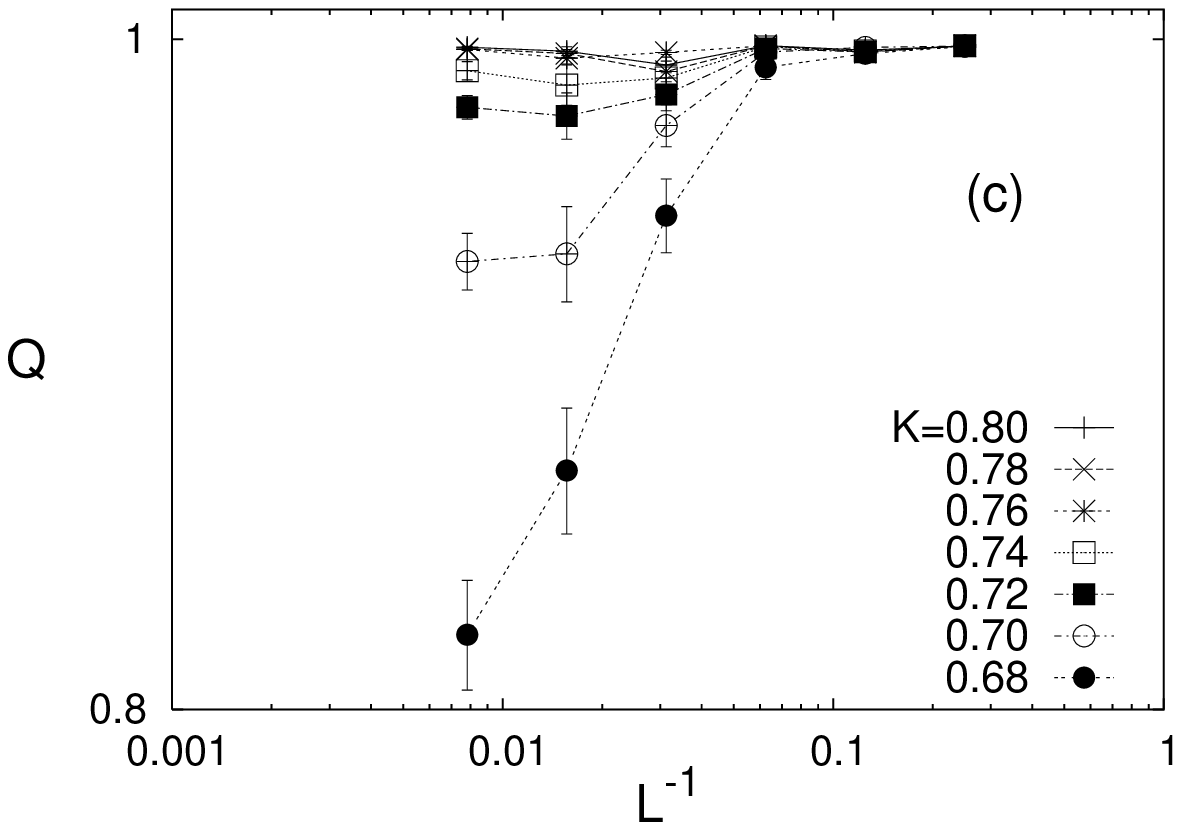}}}
\centering{\resizebox*{!}{4.2cm}{\includegraphics{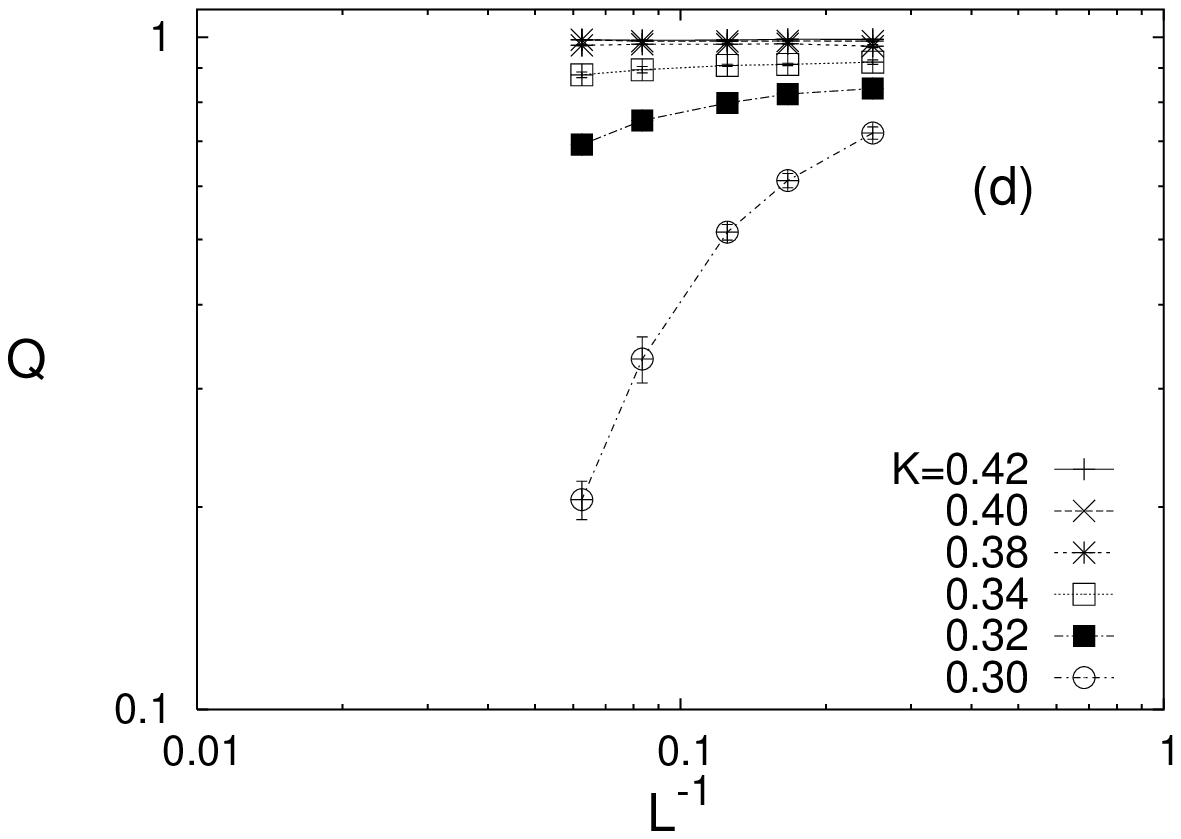}}}
\centering{\resizebox*{!}{4.2cm}{\includegraphics{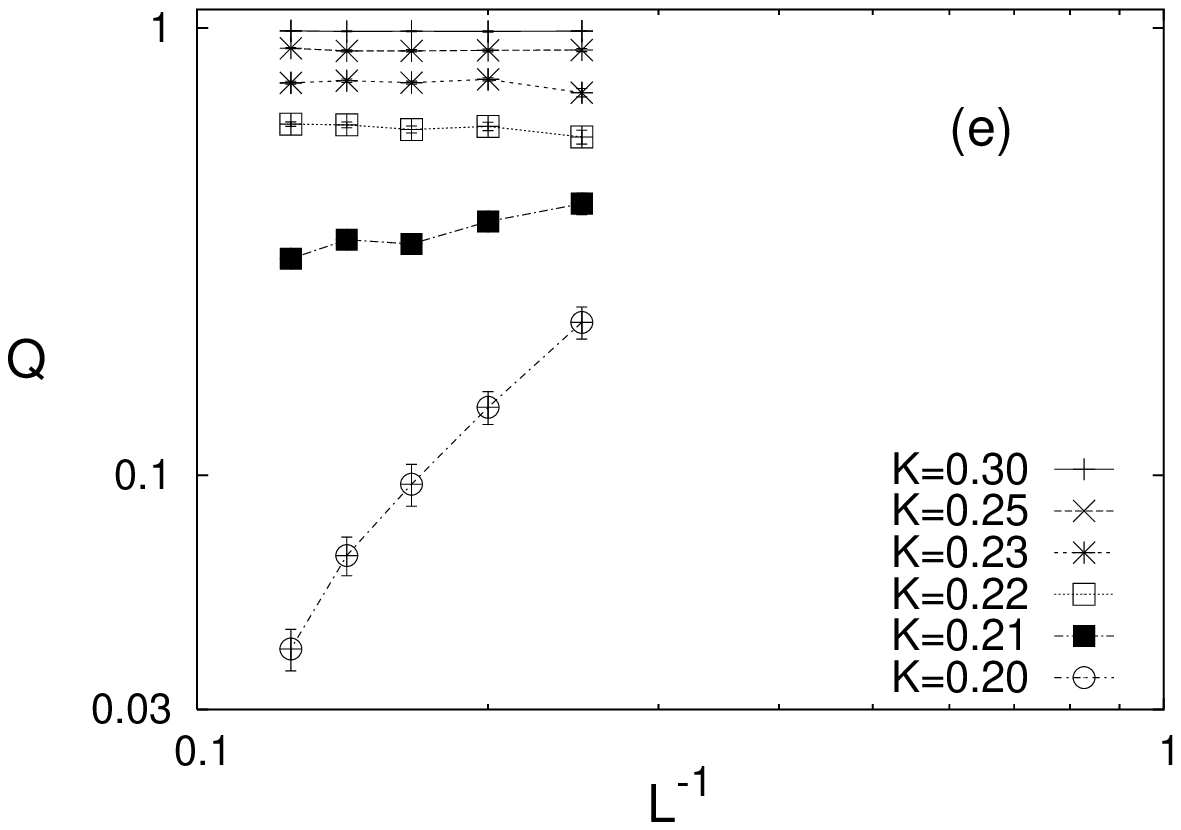}}}
\caption{Frequency order parameter $Q$ plotted as a function of the linear size 
$L$ for various values of the coupling strength $K$ and dimension $d =$ (a)\,1, (b)\,2,
(c)\,3, (d)\,4, and (e)\,5. }
\label{fig:q_L_2345D}
\end{figure}

For more quantitative analysis on the nature of the detrained phase and 
the entrainment transition, we show in Fig.~\ref{fig:q_L_2345D} the log-log plot of $Q$
versus $L^{-1}$ at various values of $K$. 
Similarly to the phase order parameter $\Delta$, we expect $Q\sim N^{-1/2}=L^{-d/2}$ 
for the fully random phase. 

Figure~\ref{fig:q_L_2345D}(a) shows that in one dimension the frequency order parameter 
decays algebraically as $Q\sim L^{-c}$, where $c$ varies very slowly from the value 
around $0.6$ at $K=3.2$ to the value $0.5$ at $K=0$. 
We presume that the apparent variation of the scaling exponent $c$ with $K$ 
reflects small-size effects and that the one-dimensional system of sufficiently large size 
exhibits the fully random phase in frequency for any finite $K$, 
with the behavior $Q\sim L^{-1/2}$. 

In two dimensions, we also face with a similar situation, except that 
$c$ varies from the value around $0.7$ at $K=1.0$ to $1.0$ at $K=0.1$.
We do not rule out the possibility to have a correlated random (detrained) phase,
similar to the three- and four-dimensional phase synchronization problems. 
However, both statistical and systematic (small-size) errors hinder us to 
clarify this point with present data. 

In higher dimensions, there is a frequency entrainment transition 
and we estimate the critical coupling strength $K_c\approx 0.71(2)$, 0.32(1), and 0.21(1) 
for $d=3$, 4, and 5, respectively. In the entrained phase for $K>K_c$, 
the order parameter $Q$ saturates to a nonzero value as $L\rightarrow\infty$, 
while in the detrained phase ($K<K_c$), it decays to zero. We examine nature 
of the detrained phase and find that it has a characteristic of the fully random phase
in five dimensions: $Q\sim L^{-2.5}$. We also investigate the critical decay, to find
$Q\sim L^{-\beta/\nu}$ with $\beta/\nu = 0.35(15)$. 
For $d=4$, we observe that $Q\sim L^{-2.0}$ in the weak-coupling regime 
(at $K\approx 0.15$), but it decays slower as $K$ is increased. However, 
the variation of the exponent is rather small (between 1.7 and 2.0), which suggests that
the entire detrained phase is also the fully random phase in four dimensions. 
The analysis at the critical point reveals that $\beta/\nu= 0.2(2)$. 
The three-dimensional situation is marginal and the present data do not provide 
even suggestive information about the nature of the detrained phase. 

It is remarkable that the phase synchronization and the frequency entrainment
transition apparently occur simultaneously (within error bars) for $d=5$, 
just like in the globally coupled system. It may be conjectured that these two
synchronization transitions always occur simultaneously for all $d\ge 5$. 
However, the critical exponents for phase synchronization and for frequency entrainment 
are clearly distinct: $(\beta/\nu)_P=1.5(3)$ and $(\beta/\nu)_F=0.35(15)$, respectively. 
Note that the two critical scalings are identical in the globally coupled system.

\begin{figure}
\centering{\resizebox*{!}{6.0cm}{\includegraphics{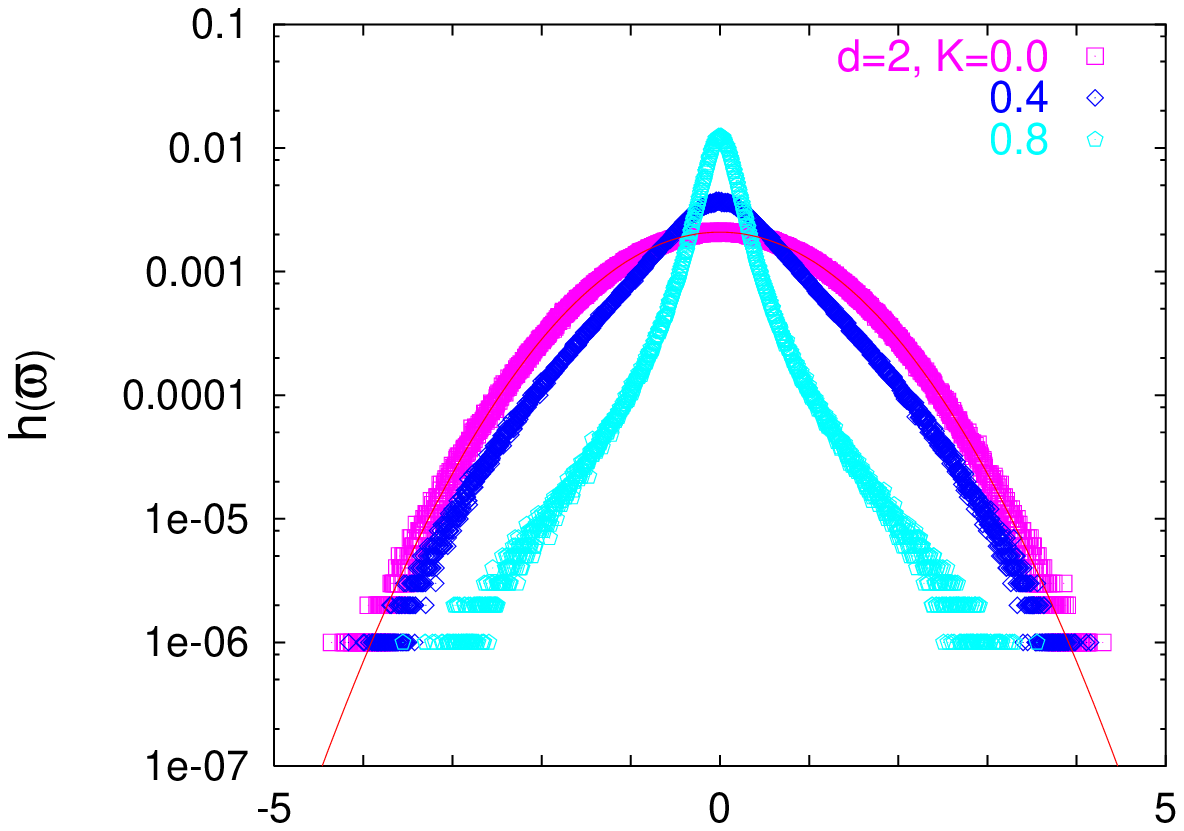}}}
\centering{\resizebox*{!}{6.0cm}{\includegraphics{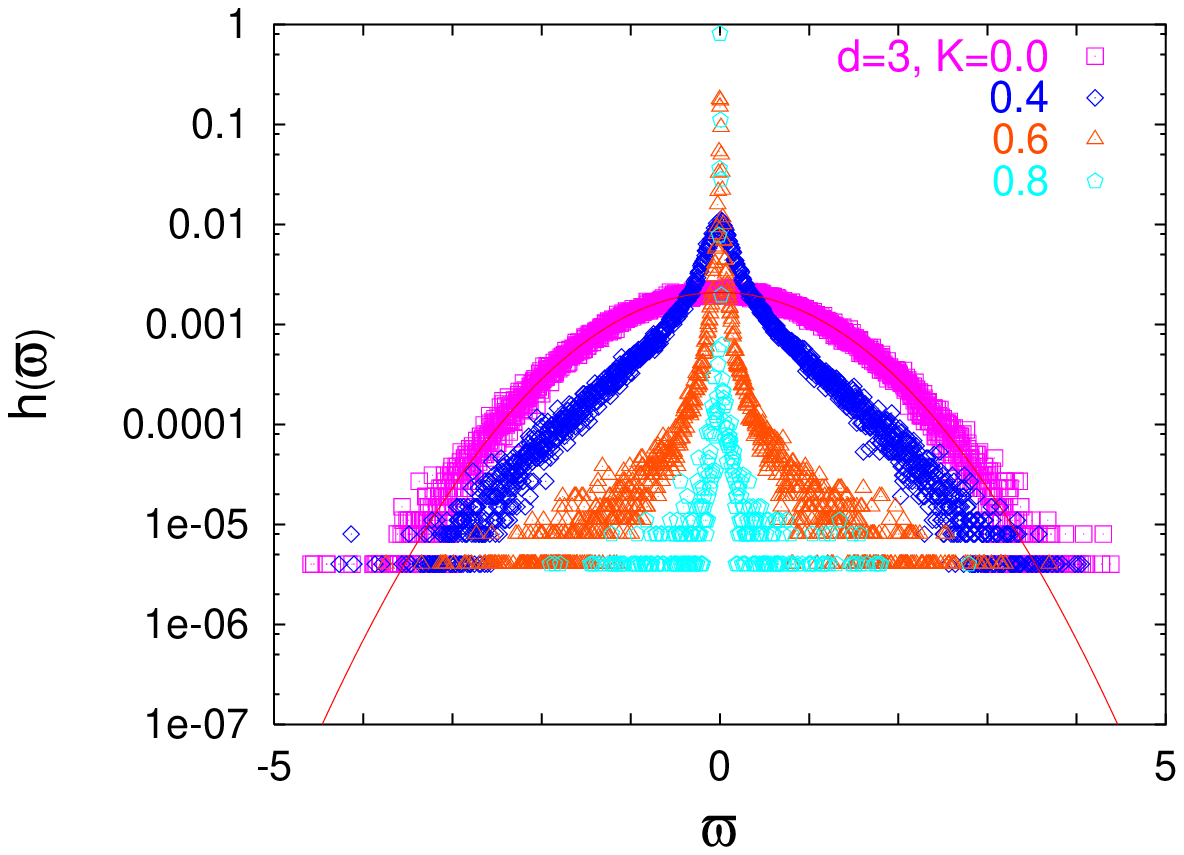}}}
\caption{Semi-log plot of the normalized histogram of the stationary-state frequency of 
oscillators for $d=2$ and 3 with $L=256$ and $64$, respectively. 
The thin curved line represents the Gaussian distribution $g({\bar\omega})$. 
}
\label{fig:PDF}
\end{figure}

Finally, we study the normalized histogram $h({\bar\omega})$ of the number of
oscillators with an identical frequency ${\bar\omega}$ in the stationary state. 
Figure \ref{fig:PDF} exhibits the semi-log plot of the histogram $h({\bar\omega})$ 
at various values of $K$ for $d=2$ and $3$ with size $L=256$ and $64$, respectively. 
At $K=0$, the histogram should be identical to the initial Gaussian distribution $g(\omega)$. 
As $K$ is increased, one can see easily that the distribution becomes narrower, 
with its peak becoming sharper. At the same time, its tail part becomes thinner 
and decays faster. 
This manifests that the oscillators having high intrinsic frequencies tend to adjust 
themselves to co-evolve with other oscillators having low intrinsic frequencies, 
through couplings between them. 
This distribution is completely distinct from that of the globally coupled system 
(see Fig.~\ref{fig:gl_hv} in the next section).

In summary, our numerical analysis shows that the frequency entrainment transition 
emerges for $d\ge 3$, which suggests that the lower critical dimension for 
frequency entrainment: $d_l^F=2$.  
For $d\ge 5$, the frequency and the phase synchronization transitions occur 
simultaneously but with different scaling behaviors.

\section{Globally coupled oscillators}

For comparison with locally coupled oscillators, we discuss some analytic and 
numerical results for globally coupled system, where each oscillator is coupled with 
every other with equal coupling strength $K/N$. The set of equations of motion reads
\begin{equation}
\frac{d\phi_i}{dt} = \omega_i - \frac{K}{N}\sum_{j=1}^{N}\sin(\phi_i - \phi_j).
\label{eq:model_gl}
\end{equation}
This system  has been much studied~\cite{ref:Kuramoto}, partly due to 
its analytical tractability. 
Similar to the system of locally-coupled oscillators, its collective behavior in phase is  
described by the order parameter $\Delta$ defined in Eq.~(\ref{eq:def_order}). 
For convenience, we define the complex order parameter according to
$\Delta e^{i\theta} \equiv N^{-1} \sum_j e^{i\phi_j}$. 
With the help of $\Delta$ and $\theta$, Eq.~(\ref{eq:model_gl}) reduces to 
$N$ identical decoupled equations
\begin{equation}
\frac{d\phi_i}{dt} = \omega_i - K\Delta\sin(\phi_i - \theta), 
\label{eq:model_gl_Delta}
\end{equation}
where $\Delta$ and $\theta$ are to be determined by imposing self-consistency. 

The stationary solution with constant $\theta$ is then obtained for the symmetric
distribution of intrinsic frequencies $g(\omega)=g(-\omega)$. A straightforward
calculation leads to the self-consistency equation~\cite{ref:Kuramoto} 
\begin{equation} \label{ordereq}
\Delta=a K\Delta - c (K\Delta)^3 + O(K\Delta)^5
\end{equation}
with $a=(\pi/2)g(0)$ and $c=-(\pi/16)g''(0)$,
which yields
$K_c = 2/\pi g(0)$~\cite{ref:Kuramoto}.
When the distribution $g(\omega)$  is concave at 
$\omega=0$, i.e., $g^{''}(0) <0$, the nontrivial solution $(\Delta\neq 0$)
appears via a pitchfork (supercritical) bifurcation as $K$ is raised
beyond $K_c$.  Near $K_c$, the nontrivial solution behaves 
as 
\begin{equation}
\Delta\propto (K-K_c)^{\beta},
\label{eq:Delta_gl}
\end{equation}
where $\beta=1/2$ is the mean-field value of the order parameter exponent.

\begin{figure}
\centering{\resizebox*{!}{6.0cm}{\includegraphics{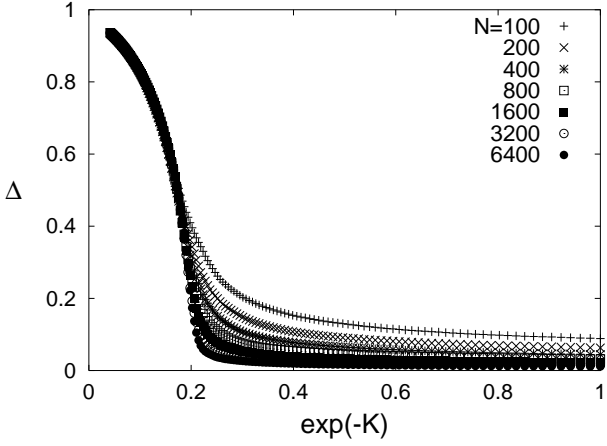}}}
\caption{Phase order parameter $\Delta$ in the globally coupled oscillator system 
shown as a function of ${\rm exp}(-K)$ for various values of $N$.}
\label{fig:gl_ph}
\end{figure}

We have also integrated numerically Eq.~(\ref{eq:model_gl}) and show in 
Fig.~\ref{fig:gl_ph} the obtained behavior of the phase order parameter $\Delta$ 
depending on ${\rm exp}(-K)$ for various values of $N$. 
As expected, there is a phase synchronization transition at $K=K_c\approx 1.6$ 
(or $e^{-K_c}\approx 0.20$), which is consistent with 
the analytic value $K_c=2/\pi g(0)=\sqrt{8/\pi}\approx 1.596$.

\begin{figure}
\centering{\resizebox*{!}{6.0cm}{\includegraphics{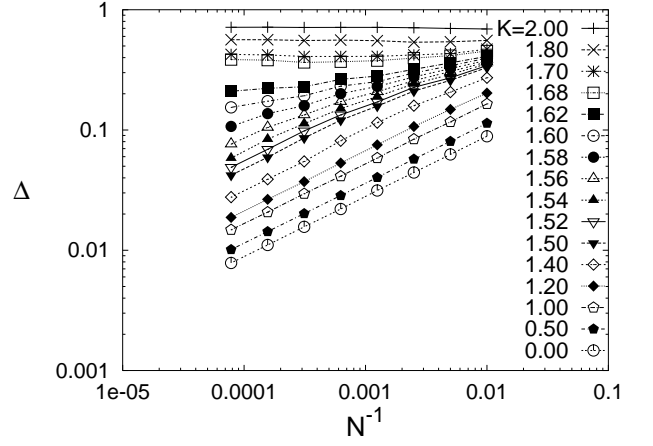}}}
\caption{Phase order parameter $\Delta$ plotted against $N^{-1}$ 
in the system of globally coupled oscillators, for various values of the coupling 
strength $K$.}
\label{fig:gl_ph_N}
\end{figure}

For more quantitative finite-size analysis, we show in Fig.~\ref{fig:gl_ph_N} 
the log-log plot of $\Delta$ versus $N^{-1}$. For $K<K_c$, we find that 
$\Delta\sim N^{-1/2}$, which implies that the desynchronized phase is the fully random phase. 
Near $K=K_c$, we assume the finite-size scaling relation 
\begin{equation} 
\Delta = N^{-\beta/{\bar\nu}} F\left[(K - K_c)N^{1/{\bar\nu}}\right],
\label{eq:scaling_gl}
\end{equation}
where the critical exponent $\bar\nu$ describes the divergence of the correlation 
volume (the number of correlated oscillators) $\xi_V$ as $K\rightarrow K_c$. 
In $d$ dimensions, we expect $\xi_{V} \sim \xi^d $ with the correlation length $\xi$,
which leads to the relation $\bar\nu = \nu d$. 
For the globally coupled system, neither the length scale nor the space dimension
have proper meaning, so only $\xi_V$ may be properly defined. 
The scaling function behaves $F(x)\sim x^\beta$ as $x\rightarrow\infty$ and 
$F(x)\sim {\rm const.}$ as $x\rightarrow 0$. 

\begin{figure}
\centering{\resizebox*{!}{6.0cm}{\includegraphics{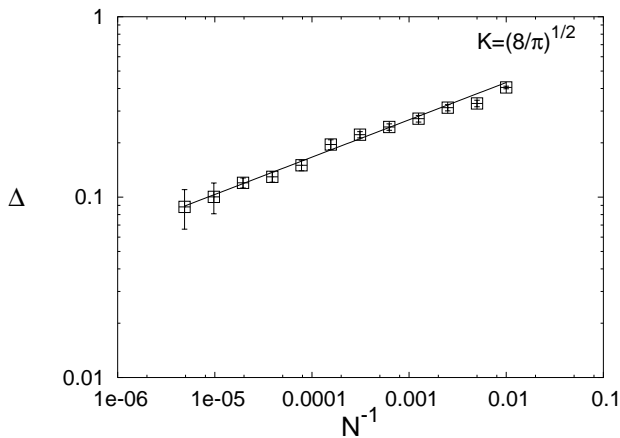}}}
\caption{Critical decay of the phase order parameter. The straight line represents
the best linear fit of slope $0.21$.}
\label{fig:gl_ph_N_Kc}
\end{figure}

At $K=K_c$, we have $\Delta\sim N^{-\beta/{\bar\nu}}$, which is shown in 
Fig.~\ref{fig:gl_ph_N_Kc}. The best fit is obtained with $\beta/{\bar\nu}= 0.21(2)$,
which, together with the exact value $\beta=1/2$, leads to the estimate ${\bar\nu}=2.4(2)$. 
Summarizing our results on the phase synchronization in the globally coupled system,
we write
\begin{equation}
\beta=1/2 \qquad {\rm and} \qquad {\bar\nu}=2.4(2).
\label{eq:nu_gl}
\end{equation}
Note that our present estimate is somewhat higher than the existing one 
$2.0(2)$~\cite{ref:Hong}. We have also checked the finite-size scaling relation directly 
by plotting $\Delta N^{\beta/{\bar\nu}}$ versus $(K/K_c -1 ) N^{1/{\bar\nu}}$ and 
found the consistent value of ${\bar\nu}=2.4(2)$. 
This may provide a hint on the upper critical dimension $d_u^P$ for phase synchronization. 
Following Ref.~\onlinecite{ref:ucd}, one may assume that the relation 
${\bar\nu}=\nu_{\rm MF}~ d_u^P$ holds in the coupled synchronization problem. 
Then, with the usual mean field value $\nu_{\rm MF}=1/2$, we come up with $d_u^P\approx 5$. 
If this turns out to be correct, our result for the locally coupled system in five
dimensions may reflect the log-type corrections to the mean-field values. 
At present, however, it is too early to conclude on the value of the upper critical dimension.

\begin{figure}
\centering{\resizebox*{!}{6.0cm}{\includegraphics{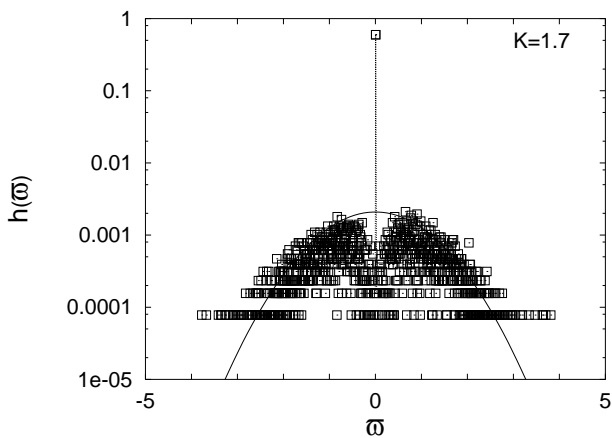}}}
\caption{Semi-log plot of the normalized histogram of the stationary-state frequencies 
of oscillators in the entrained phase ($K=1.7$). 
The thin curved line represents the Gaussian distribution $G({\bar\omega})$}
\label{fig:gl_hv}
\end{figure}

We now study the frequency entrainment behavior for the globally coupled oscillator system.
Similarly to the case of the locally-coupled system, we investigate
the normalized histogram $h({\bar\omega})$ of the number of oscillators with
an identical frequency ${\bar\omega}$ in the stationary state.
Figure~\ref{fig:gl_hv} shows the histogram obtained numerically in the
frequency entrained phase ($K>K_c$), which exhibits a delta-function peak
at the entrained frequency $({\bar\omega}=0$) in the background Gaussian-type 
distribution. Looking carefully at the background distribution, we find two 
symmetric humps near ${\bar\omega}=0$ and its Gaussian shape for large $|{\bar\omega}|$
is almost identical to the initial frequency distribution 
$g({\bar\omega})$~\cite{ref:Kuramoto,ref:Sakaguchi}. 
On the other hand, in the detrained phase ($K <K_c$), there are neither peak nor humps
and the histogram $h({\bar\omega})$ is identical to $g({\bar\omega})$.
This implies that the detrained phase for the globally coupled system is fully random.
(Of course, in a finite-size system, the frequency order parameter is nonzero at all $K$,
so one can see the peak-hump structure even for $K<K_c$. However, in the thermodynamic
limit, the peak-hump structure should disappear for $K<K_c$.)

As $K$ increases beyond $K_c$, the oscillators with very low intrinsic frequencies 
($\omega_i\approx 0$) are entrained first, so the histogram shows two humps and dips near the
entrained frequency (${\bar\omega}=0$). This distribution is clearly distinct from 
that of the locally coupled system, where the oscillators with high intrinsic frequencies 
are affected first due to local coupling with nearby oscillators with low intrinsic frequencies 
(see Fig.~\ref{fig:PDF} and Sec.~IV). In the globally coupled system, 
all oscillators are coupled with the same strength (no local environment involved), 
and accordingly oscillators with similar frequencies are entrained first. 

The analytic form of the normalized histogram in the thermodynamic limit is given 
by~\cite{ref:Kuramoto,ref:Sakaguchi}
\begin{equation}
h(\bar\omega)=Q\delta(\bar\omega) 
+\frac{|\bar\omega|}{\sqrt{{\bar\omega}^2+(K\Delta/\bar\omega)^2}}
g\left(\sqrt{{\bar\omega}^2+(K\Delta/\bar\omega)^2}\right),
\label{eq:dist_anal}
\end{equation} 
where the frequency order parameter $Q$ is determined by the normalization condition.
A concise expression for $Q$ then reads
\begin{equation}
Q =\int_{-K\Delta}^{K\Delta}g(\omega) d\omega.
\label{eq:Q_re_gl}
\end{equation}

For $\Delta=0$, one can easily see that $h(\bar\omega)=g(\bar\omega)$. 
On the other hand, for $\Delta>0$, a delta-function peak and two humps appear in 
Eq.~(\ref{eq:dist_anal}), as expected.
For small $\Delta$ (near criticality), the frequency order parameter behaves as
\begin{equation}
Q\simeq 2K\Delta g(0)\propto (K-K_c)^{\beta} ~~ {\rm with} ~~ \beta=1/2.
\label{eq:Q_gl}
\end{equation}
Note that both phase synchronization and frequency entrainment share 
the same critical point and the same scaling behavior near criticality.

\begin{figure}
\centering{\resizebox*{!}{6.0cm}{\includegraphics{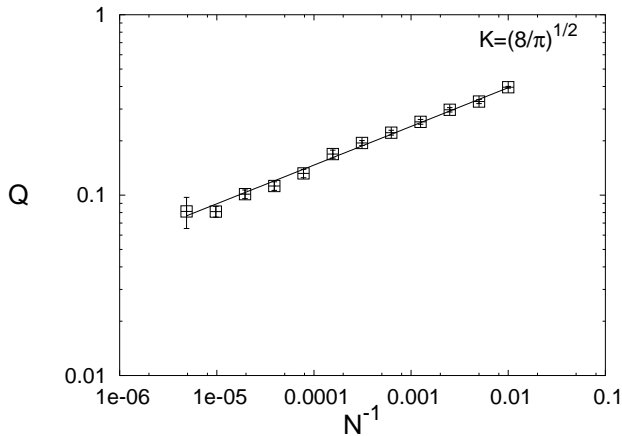}}}
\caption{Critical decay of the frequency order parameter $Q$. The straight line represents
the best linear fit of slope $0.21$.}
\label{fig:gl_Q_N_Kc}
\end{figure}

We also study the finite-size scaling behavior of $Q$ at criticality and
show the log-log plot of $Q$ versus $N^{-1}$ in Fig.~\ref{fig:gl_Q_N_Kc}.
It is observed that $Q\sim N^{-\beta/{\bar\nu}}$ with $\beta/{\bar\nu}=0.21(2)$,
which is also consistent with the value for the phase order parameter $\Delta$.
In the detrained phase, we find $Q\sim N^{-1/2}$, as expected.

\begin{figure}
\centering{\resizebox*{!}{6.0cm}{\includegraphics{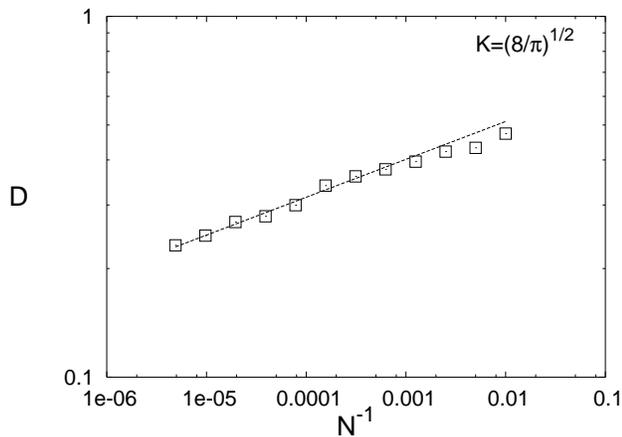}}}
\caption{Critical decay of the frequency order parameter $D$. The straight line represents
the best linear fit of slope $0.105$.}
\label{fig:gl_D_N_Kc}
\end{figure}

As the humps disappear as $K\rightarrow K_c^+$, one may define the frequency distance $D$
between the delta peak position (${\bar\omega}=0$) and the hump position,
as another frequency order parameter. From Eq. (\ref{eq:dist_anal}), it is straightforward
to show $D\sim(K\Delta)^{1/2}\sim Q^{1/2}$. One thus expects $D\sim N^{-\beta/(2{\bar\nu})}$ 
at criticality and $D\sim N^{-1/4}$ in the detrained phase. 
We estimate from Fig.~\ref{fig:gl_D_N_Kc} that $\beta/(2{\bar\nu})=0.105(10)$, 
which is also fully consistent with the estimate from $Q$. 

Finally, we comment on the possibility of the complete frequency entrainment 
$(Q=1)$ in the system of globally coupled oscillators. From Eq. (\ref{eq:Q_re_gl}), 
one can easily notice that, if $g(\omega)$ has no cutoff at finite frequency $\omega$
like our choice $g(\omega)\sim \exp(-\omega^2/4\sigma)$, $K$ should grow arbitrarily 
large to yield $Q=1$.  However, with finite cutoffs in $g(\omega)$, the system can 
exhibit complete frequency entrainment at finite values of $K$. 

\section{Summary}
We have explored the collective behavior of locally coupled oscillators
with random intrinsic frequencies on $d$-dimensional hypercubic lattices
as well as globally coupled oscillators where each oscillator is coupled with
every other one with the same strength.
Both phase synchronization and frequency entrainment have been studied. 
We have probed those phenomena through one typical model of growing surface.  
By measuring the mean-square width for both the phase and the phase velocity 
of the growing interface, we have estimated some linearity characteristics of the system. 
In particular, the lower critical dimension for phase synchronization 
has been obtained to be four.  
For frequency entrainment, the lower critical 
dimension has been numerically found to be two. 
The effects of the nonlinear sine coupling on the phase synchronization
as well as the frequency entrainment have also been 
investigated by means of numerical simulations; 
revealed is that the sine coupling tends to suppress not only phase synchronization 
but also frequency entrainment rather than enhancing those. 
It is found that phase synchronization emerges for $d>4$ and frequency entrainment 
transition occurs for $d>2$. 
The nature of the phase synchronization transition in five dimensions has also been 
explored and the critical exponents $\beta$ and $\nu$ have been measured. 
The values are observed to differ from those for the globally coupled system, 
manifesting that the five-dimensional system belongs to the different universality 
class from the mean-field one. It further allows the speculation that the 
upper critical dimension for phase synchronization might be higher than five, 
although further study is necessary for conclusive results. 

\section*{Acknowledgments}
One of us (MYC) was supported in part by the KOSEF Grant No. R01-2002-000-00285-0
and by the BK21 Program.

\end{document}